%% file: main.tex
\documentclass[a4paper,11pt]{article}
\pdfoutput=1
\usepackage{jcappub}

\usepackage{shortcuts}
\graphicspath{{./figs/}}

\begin{document}
\title{Impact of Large-Scale Structure Systematics on Cosmological Parameter Estimation}

\author[a]{Humna Awan\orcidlink{0000-0003-2296-7717},}
\affiliation[a]{Leinweber Center for Theoretical Physics, Department of Physics, University of Michigan, Ann Arbor, MI 48109, USA}
\author[b]{Eric Gawiser\orcidlink{0000-0003-1530-8713},}
\affiliation[b]{Department of Physics $\&$ Astronomy, Rutgers University, 136 Frelinghuysen Rd., Piscataway, NJ 08554, USA}
\author[c]{Javier Sanchez\orcidlink{0000-0003-3136-9532},}
\affiliation[c]{Space Telescope Science Institute, 3700 San Martin Dr, Baltimore, MD 21218, USA}
\author[d]{Ignacio Sevilla-Noarbe\orcidlink{0000-0002-1831-1953},}
\affiliation[d]{Centro de Investigaciones Energ\'eticas Medioambientales y Tecnol\'ogicas (CIEMAT), Av. Complutense 40, 28040 Madrid, Spain}
\author{the LSST  Dark Energy Science Collaboration}

\emailAdd{hawan@umich.edu}

\abstract{
Large near-future galaxy surveys offer sufficient statistical power to make our cosmology analyses data-driven, limited primarily by systematic errors. Understanding the impact of systematics is therefore critical. We perform an end-to-end analysis to investigate the impact of some of the systematics that affect large-scale structure studies by doing an inference analysis using simulated density maps with various systematics; these include systematics caused by photometric redshifts (\pz s), Galactic dust, structure induced by the telescope observing strategy and observing conditions, and incomplete covariance matrices. Specifically, we consider the impacts of incorrect \pz\ distributions (photometric biases, scatter, outliers; spectroscopic calibration biases), dust map resolution, incorrect dust law, selecting none or only some contaminant templates for deprojection, and using a diagonal covariance matrix instead of a full one. We quantify the biases induced by these systematics on cosmological parameter estimation using tomographic galaxy angular power spectra, with a focus on identifying whether the maximum plausible level of each systematic has an adverse impact on the estimation of key cosmological parameters from a galaxy clustering analysis with Rubin Observatory Legacy Survey of Space and Time (LSST). We find \pz\ systematics to be the most pressing out of the systematics investigated, with spectroscopic calibration biases leading to the greatest adverse impact while helpfully being flagged by a high $\chi^2$ value for the best fit model. Larger-than-expected \pz\ scatter, on the other hand, has a significant impact without necessarily indicating a poor fit. In contrast, in the analysis framework used in this work, biases from observational systematics and incomplete covariance matrices are comfortably subdominant.
}

\keywords{cosmological parameters from LSS, galaxy clustering}

\maketitle
\section{Introduction\label{sec: intro}}
The field of observational cosmology is undergoing a rapid transition with the advent of large galaxy surveys -- surveys that will provide us with unprecedented amounts of data, greatly reducing statistical limitations. In this new era, it is imperative to quantify the impact of systematic uncertainties that exceed the statistical uncertainties and to mitigate them.

Large-scale structure (LSS) as a cosmological probe is no exception. LSS can be traced by the angular clustering in the spatial distribution of galaxies (see e.g., \citep{cooray+2002}) using power spectra or two-point correlation functions (2PCFs), and current generation surveys have yielded some of the tightest cosmology constraints from these tracers (see e.g., \citet{des-3x2pt-y3} for Dark Energy Survey results; \citet{hsc-y3-1}, \citet{hsc-y3-2} for those from Hyper Suprime-Cam; and \citet{kids} for results from the Kilo-Degree Survey). Forthcoming large galaxy surveys will provide us with at least an order of magnitude more galaxies than present surveys, making the shot noise subdominant to sample variance caused by inherent randomness of the initial density field and to systematic uncertainties.  For our analyses to achieve the promised accuracy, we will need to operate in the regime where systematic uncertainties are subdominant. New 2PCF estimators that account for various systematics (e.g., \pz\ uncertainties as in \citealt{ross+2017}, \citealt{awan+2020}; fibre assignments as in \citealt{bianchi+2018}) as well as other contaminant mitigation methods (e.g., those presented in \citealt{shafer+2015}, \citealt{elsner+2016}, \citealt{awan+2016}) are active areas of development.

Given the changing landscape, investigations into the impact of various systematics are an active area of work. This includes systematics studied here, i.e., \pz\ uncertainties (see e.g., \citet{wright+2020} for improved \pz\ distribution estimation for better cosmology inference), impacts of Milky Way dust (see e.g., \citet{chiang+2019}, probing LSS imprints on dust maps), as well as other observational contaminants (references above as well as e.g., \citet{johnston+2021, lochner+2022}). Going beyond galaxy clustering analyses, there are developments in understanding the impacts of systematics playing a role in joint analyses, such as with cosmic shear (see e.g., \citet{rodriguez+2022} and references therein), also entailing probing systematics affecting shear specifically (e.g., \citet{campos+2023, leonard+2024}).

In this work, we focus on some of the systematics that impact LSS studies for cosmological parameter estimation with photometric galaxy surveys, with the goal to identify the impact of the maximum plausible level of given systematics on cosmological parameter estimation. We perform this analysis in the context of \href{https://www.lsst.org/}{Vera C. Rubin Observatory Legacy Survey of Space and Time} (LSST; \citealt{abell+2009}, hereafter referred to as the LSST Science Book). Specifically, the LSST Dark Energy Science Collaboration (DESC) is planning a tomographic galaxy clustering analysis to constrain key cosmological features including dark energy; the details of the planned analyses can be found in the DESC Science Requirements Document (\citealt{desc-srd2018}; hereafter referred to as the DESC SRD). While our work is applicable beyond LSST/DESC, where we need specifics, we use those laid out for DESC analysis of the LSST data (as detailed in \citetalias{desc-srd2018}).

The structure of this paper is as follows: in \autoref{sec: methodology}, we explain our workflow and analysis details. In \autoref{sec: systematics}, we lay out the systematics considered, while \autoref{sec: results} shows the results. We conclude and discuss future directions in \autoref{sec: conclude}.

\section{Methodology\label{sec: methodology}}
\begin{figure}[!htb]
	\centering
	\includegraphics[width=\linewidth, trim={10 0 70 0}, clip=true]{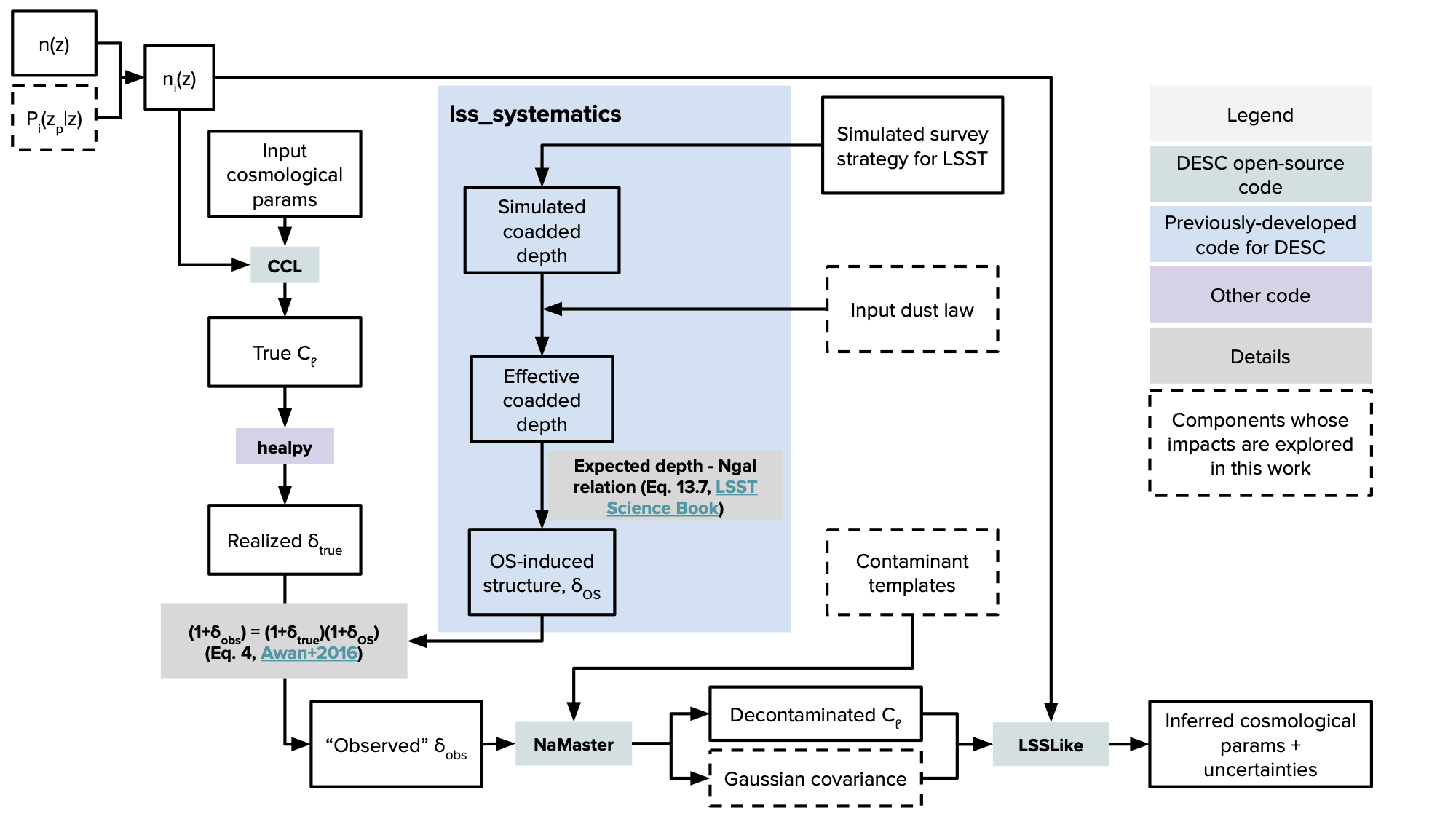}
	\caption{Workflow for our analysis pipeline. The text elaborates on how we set up the photometric redshifts and galaxy number distribution ($\mathcal{P}_i(z_p | z)$ and $n_i(z)$ respectively; \autoref{sec: pz-nz-model}), which are used to realize true galaxy density maps ($\delta_\mathrm{true}$; \autoref{sec: map generation}) and then the observed ones ($\delta_\mathrm{obs}$; \autoref{sec: observed maps}), leading us to observed galaxy power spectra (\autoref{sec: estimation of cls}) and then the posteriors (\autoref{sec: estimation of params}). We also elaborate on how we test the impact of components like $\mathcal{P}_i(z_p | z)$ (\autoref{sec: pz models}), dust law (\autoref{sec: mw models}), contaminant templates (\autoref{sec: ct}), and the covariance matrix (\autoref{sec: cov}).}
	\label{fig: workflow}
\end{figure}

In order to test the impacts of various systematics on LSS studies, we perform an end-to-end analysis using simulated galaxy density maps, with the overall workflow of the analysis pipeline shown in \autoref{fig: workflow}. To summarize, given a fixed cosmology, we first generate theoretical galaxy power spectra which are combined with observational artifacts to generate observed galaxy density maps. Then, as we would for real observed density maps, we measure the galaxy power spectra while accounting for various contaminants and perform a likelihood analysis for cosmological parameters. We discuss each of these steps in detail in the following subsections; the workflow shows the various steps.

We carry out the analysis as planned by the DESC (as outlined in \citetalias{desc-srd2018}), i.e., for ten tomographic bins, spanning redshift $z = 0.2-1.2$, with constant redshift bin width of $\Delta z=0.1$; the lever arm between low-$z$ and high-$z$ data is particularly important for constraints from features that depend on coordinate distance, like the Baryonic Acoustic Oscillations.

\subsection{Photometric Redshifts and Galaxy Number Distribution\label{sec: pz-nz-model}}
Before we generate power spectra, we construct models for photometric redshifts and the galaxy number distribution. To achieve this, we loosely follow Section 13.3.3 in \citetalias{abell+2009} to construct the galaxy redshift distribution for each observed redshift bin $i$ while accounting for \pz\ errors, with specifics based on \citetalias{desc-srd2018} where possible. Assuming a given photometric redshift error distribution $\mathcal{P}(z_p; z)$, we can construct the probability of assigning a galaxy to the photometric redshift bin $i$, with bin edges $\zpmin, \zpmax$, as follows:
\eq{
	\mathcal{P}_i(z)
	\propto
		\displaystyle\int_{\zpmin}^{\zpmax}
		dz_p' \ \mathcal{P}(z_p'; z)
	\label{eq: joint pz general}
}
Furthermore, Equation 13.10 in \citetalias{abell+2009} defines the underlying galaxy redshift distribution as
\eq{
	n(z) = c_\mathrm{norm} z^\alpha \mathrm{exp}\bsqbr{ - \bbr{ \frac{z}{z*} }^\beta}
	\label{eq: nz}
}
where $\alpha=2, z*=0.28$, $\beta=0.9$, as specified in \citetalias{desc-srd2018}; $c_\mathrm{norm}$ is the normalization constant that would yield the number of galaxies for a given redshift. In order to get the normalization, we consider Equation 3.7 in \citetalias{abell+2009}, which gives us the galaxy number density for $20.5 < i < 25.5$:
\eqs{
	\eta_\mathrm{gal}
	&= 46 \times 3600 \times 10^{0.31(i-25)} \mathrm{galaxies / deg^2}
	\label{eq: ngal-lsst}
}
where $i$ denotes the dust-corrected, extended-source limiting magnitude in $i$-band. Note that given the exponential form, $N_\mathrm{gal} \propto \eta_\mathrm{gal}$.

\noindent Now, following e.g., \citealt{ma+2006}, we have
\eq{
	n(z) = \frac{d}{dz} \frac{dN_\mathrm{gal}}{d\Omega} = \frac{d}{dz}\eta_\mathrm{gal}
	\label{eq: nz-Nz}
}
which, in combination with \autoref{eq: nz} and \autoref{eq: ngal-lsst}, gives the normalization constant, $c_\mathrm{norm}$. Finally, we can write the true redshift distribution of galaxies that get placed in photometric redshift bin $i$ as
\eq{
	n_i(z) = \mathcal{P}_i(z) n(z)
	\label{eq: nz_i pz}
}
Since the next practical step in \pz\ fitting is spectroscopic calibration via cross-correlation with a spectroscopic galaxy sample, we follow \citet{cawthon+2022} in choosing a 2-parameter model to address the impact of errors in this calibration: shift parameter, $\Delta_i$, which shifts the $n_i(z)$ for a given redshift bin $i$ (with baseline value being 0), and stretch parameter, $s_i$, which stretches a given $n_i(z)$ as if it was centered around the mean (shifted) redshift of galaxies in the bin (with the baseline value being 1). Therefore, we have
\eq{
n_i(z) \rightarrow n_i(z') - \Delta_i
\label{eq: nz_i delta}
}
which, in implementation, means that $z$ becomes $z+\Delta_i$, and
\eq{
n_i(z) \rightarrow \frac{1}{s_i} n_i(z'), \ \ \ z' =  \frac{(z-z_{\mathrm{mean}})}{s_i} +z_{\mathrm{mean}}\
\label{eq: nz_i s}
}
which, in implementation, entails $z$ becoming $(z  - z_{\mathrm{mean}})  s_i  + z_{\mathrm{mean}}$,
where $z_\mathrm{mean}$ is the mean redshift in each bin.\footnote{Practical notes: when implementing in code, we shift and stretch, in that order, the input $z$ and $n_i(z)$, and discard values for negative $z$; the modified redshift arrays have the same spacing as the original but the length may not be the same depending on how much shift/stretch is implemented. Also, the mean redshift is calculated as the weighted mean for the given redshift bin, i.e., $\sum z_i n_i(z) / \sum n_i(z)$.} With this formulation in hand for $n_i(z)$, we can move on to generating theoretical galaxy power spectra.

\subsection{True Galaxy Power Spectra and Density Maps\label{sec: map generation}}
Taking the galaxy number distribution in each of our redshift bins, $n_i(z)$, alongside fiducial cosmological parameters and a formulation for galaxy bias, we run Core Cosmology Library (\ccl) to generate galaxy power spectra; implementation details of \ccl\ can be found in \citet{ccl}. For our fiducial cosmology, as passed to \ccl, we assume $\Omega_c = 0.27$; $\Omega_b = 0.045$; $\Omega_k, \Omega_\nu = 0.0$; $h_0 = 0.67$; $\sigma_8 = 0.8$; $n_s = 0.96$; $w_0 = -1$; $w_a = 0$.
We use the transfer function given by \citet{eisenstein+hu1998}, and neglect non-linear power spectrum contributions, redshift-space distortions and magnification effects. For galaxy bias, we assume $b(z) = 1 + 0.84z$, following \citetalias{abell+2009} (Section 13.3.2).

With theory galaxy spectra in hand, we use \ttt{healpy}\footnote{\url{http://healpix.sf.net}} routine \ttt{synfast} to generate the galaxy density maps for each of the redshift bins. In order to preserve the cross-correlations within the framework of \ttt{synfast}, which gives every realized mode a random amplitude, we follow \citet[Appendix A]{giannantonio+2008} to create final maps using random phases that preserve the cross correlations. Note that since we use $n_i(z)$ calculated while accounting for photometric redshift errors, these maps effectively have photometric redshift contamination. For our maps, we use \healpix\ resolution given by $N_\mathrm{side} = 1024$\footnote{We realize this is lower resolution than typical (e.g., \nside\ 4096 as in \citealt{rodriguez+2022}) but since we are not using image-based maps, we choose speed over trying to capture small-scale structure that is not present in our maps.}, which divides the sky into $1.26 \times 10^7$ equal-area pixels of area $11.8$ arcmin$^2$.

\subsection{Observed Galaxy Density Maps\label{sec: observed maps}}
In order to simulate the observed galaxy density maps, we must account for observational artifacts, for which we follow Equation 4 in \citet{awan+2016}:
\eq{
	\bbr{1 + \delta_{\mathrm{observed}, i}} = \bbr{1+\delta_{\mathrm{true}, i}}\bbr{1+\delta_{\mathrm{OS}}}
}
where $\delta_{\mathrm{OS}}$ is the artificial LSS induced by various observational systematics, including the observing strategy and Milky Way dust. Note that we do not add shot noise since it should be subdominant for a Year 10 LSST sample; we do test the impact of the random seed used to generate the simulated maps as discussed in \autoref{sec: theory-seeds}.

\subsubsection{Observational Artifacts Considered\label{sec: del-os}}
In order to simulate observational artifacts, we consider a simulation that contains the relevant details of a 10-year, simulated LSST survey. An overview of the simulations can be found in \citet{connolly+2014} while the specifics of the scheduler can be found in \citet{naghib+2019}; we use one of the recent baseline simulations, \ttt{baseline\_nexp2\_v1.7\_10yrs.db}, details of which can be found in \citet{jones+2021}. We access and process the simulation using LSST Metrics Analysis Framework (MAF; \citet{jones+2014})\footnote{\url{https://rubin-sim.lsst.io/maf.html
}}, designed to allow an easier analysis of the simulated survey data.

To simulate the artifacts, we utilize the code pipeline set up for \citet{awan+2016} which allows us to construct the density fluctuations induced by the observational systematics. Specifically, following \citet{awan+2016}, we use \autoref{eq: ngal-lsst} to covert the 5$\sigma$ coadded depth ($5\sigma_{\mathrm{stack},j}$) into an estimated number of galaxies for each pixel $j$ \citep[following the prescription in][Eq. 2]{awan+2016}. We start with the galaxy number density:
\eqs{
\hspace*{-1.5em}
	\eta_{\mathrm{gal}, j}
    &= \frac{1}{2} \int_{-\infty}^{i_\mathrm{lim}} 
		dm \ {
            \mathrm{erfc}(m-5\sigma_{\mathrm{stack},j} )
            \bbr{
                \frac{d}{dm'} \eta_\mathrm{gal}(m') 
                }\Bigg|_{m=m'}
        }
	\label{eq: ngal-erfc}
}
where $\eta_\mathrm{gal}(m')$ is given by \autoref{eq: ngal-lsst}. Here, the \texttt{erfc} function accounts for incompleteness\footnote{As discussed in \citet{awan+2016}, the modeling choice of an \ttt{erfc} to account for incompleteness ensures that the completeness remains damped for higher magnitudes as opposed to e.g., the Fleming function \citep{fleming+1995} with which the completeness rises after $r \sim 30$.} in the 5$\sigma$ depth for extended source detections in pixel $j$, and $i_{\mathrm{lim}}$ specifies the 
extended source $i$-band magnitude limit (=25.2, chosen to deliver the gold sample). To get the number of galaxies in each pixel $j$, we scale the quantity calculated via \autoref{eq: ngal-erfc} by the area of the pixel. Note that this treatment is in contrast to \citet{awan+2016} which used the analogs of \autoref{eq: ngal-lsst} from mocks to feed into \autoref{eq: ngal-erfc}, given the focus on analyzing the artifacts induced for specific redshift bins (instead of an overall case, as being done here).

Furthermore, going beyond \citet{awan+2016}, we restrict our analysis to the survey footprint that yields the S/N goals needed for dark energy measurements with LSST as detailed in \citetalias{desc-srd2018}; this is achieved, following \citealt{lochner+2018}, by using a depth cut ($i > 25.9$ where $i$ is point-source dust-extinguished, coadded $i$-band depth after the ten-year survey\footnote{This cut on point-source depth is roughly equivalent to an extended source depth cut of $i>25.2$.}), a dust extinction cut ($E(B-V) < 0.2$), and requiring coverage in all six LSST filters.

\subsection{Estimation of Galaxy Power Spectra\label{sec: estimation of cls}}
We take the observed galaxy density maps generated following \autoref{sec: observed maps} and pass them to \namaster\ which implements template deprojection to remove the contaminants affecting the observed galaxy density\footnote{This is done by assuming that $\delta_\mathrm{obs} = \delta_\mathrm{true} + \sum_k^{N_t} \alpha_k f^k$, where $ \alpha_k$ is the unknown amplitude of the $k$th contaminant map $f^k$, with $N_t$ total contaminant maps. Template deprojection then essentially projects the observed density maps onto the space orthogonal to the (assumed-linear) contaminants' templates; see more in \citet{elsner+2016, alonso+2019}.}; details about \nmt\ can be found in \citet{alonso+2019}. We also provide \nmt\ the skymaps for various contaminants (i.e., observational artifacts); these are discussed further in \autoref{sec: ct}. With these inputs, \nmt\ estimates the angular power spectra (with mode coupling given not-full-sky maps; and windowing) as well as the Gaussian covariances; we use $\ell_\mathrm{max}$ = 850, with band powers constructed using 20 log-spaced $\ell$-samples between 5 and 850. 

We then implement $\ell_\mathrm{max}$ clipping, discarding multipoles for each redshift bin that probe scales smaller than $k_\mathrm{max}$ = 0.2; this clipping is done using comoving distance calculations for the fiducial cosmology.  This leads to an 
$\ell_\mathrm{max}$ = \{209.58, 285.83, 357.73, 425.42, 489.09, 548.99, 605.34, 658.39, 708.37, 755.52\}, respectively for the ten redshift bins, leading us to keep \{15, 16, 17, 17, 18, 18, 19, 19, 19, 20\} out of the 20 multipole bins respectively.

To interface with the likelihood module, we save the \nmt\ output (spectra and covariance matrix) as a \sacc\ object\footnote{\sacc\ is a file formatting DESC package to Save All Correlations and Covariances; \url{https://github.com/LSSTDESC/sacc}}.

\subsection{Likelihood of Cosmological Parameters\label{sec: estimation of params}}
Using the power spectra, we start with constraining five parameters: $\theta = \{\Omega_M h^2, w_0, w_a, b_1^{\mathrm{linear}},\allowbreak b_N^{\mathrm{linear}}\}$, where we consider the first three as cosmological parameters and the latter two as nuisance parameters\footnote{Since we fix $\sigma_8$ and constrain $\Omega_M$, we are treating galaxy bias parameters as nuisance parameters.  Constraining cosmology using galaxy bias is an active area of R\&D, and exploration of systematics impacts on these constraints is left for future work.}. Specifically, $\Omega_M$ is the matter density; $h$ is the reduced Hubble constant; $w_0, w_a$ specify the dark energy (DE) equation of state; while $b_1^{\mathrm{linear}}, b_N^{\mathrm{linear}}$ are the bias parameters in the first and the last redshift bin ($N$=10 here).  The superscript signifies that our setup assumes a linear evolution of galaxy bias versus redshift given by $b(z_i) = b_1^{\mathrm{linear}} + (b_N^{\mathrm{linear}} - b_1^{\mathrm{linear}}) (z_i - z_1) / (z_N - z_1)$. We also add a shift and stretch parameter for each redshift bin ($\Delta_i$, $s_i$), following \citealt{cawthon+2022}. This means that for the 10-bin case, we have a total of 25 parameters to fit. We choose not to vary $n_s$ or $H_0$, given weak expected constraints; and $\sigma_8$, given the use of linear theory\footnote{For simplicity, we fix these parameters; in a full analysis, these would be fit but then marginalized over.} as well as the parameter's degeneracy with galaxy bias, which cannot be broken in a galaxy clustering only analysis.

To get the predicted spectra for any parameter set $\theta$ in the parameter space, we construct an \ttt{LSSTheory} object using the theory \sacc\ object. This allows us to easily recycle the tracers' information alongside the $\ell$-sampling when generating spectra corresponding to $\theta$ using \ccl, using the same theory used to generate the true spectra, as explained in \autoref{sec: estimation of cls}. We set up flat priors while using physical constraints on the parameters where applicable; see details in \autoref{tab: priors}.

\begin{table*}[!htb]
\resizebox{\linewidth}{!}{
\centering
\begin{tabular}{c c c}
\hline
Parameters & Prior & Description \\
\hline
$\Omega_M h^2$ & $\mathcal{U}(\Omega_b h_0^2, 0.4)$ & total matter density fraction multiplied by the reduced Hubble constant
\\
$w_0$ & $\mathcal{U}(-2.5, 0.5)$ & DE equation of state parameter; specifies state at $z=0$  \\
$w_a$ & $\mathcal{U}(-1.5, 1.5)$ & DE equation of state parameter; specifies redshift evolution \\
$b_j^{\mathrm{linear}}$ & $\mathcal{U}(0, 3)$ &  galaxy bias for $j=1,10$ redshift bins; specify linear galaxy bias  \\
$\Delta_i$ & $\mathcal{U}(-0.2, 0.2)$ & \pz\ stretch parameter for $i$th redshift bin \\
$s_i$ & $\mathcal{U}(0.5, 2)$ & \pz\ shift parameter for $i$th redshift bin \\
\hline
\end{tabular}
}
\caption{Descriptions of parameter fit and their priors; totaling 1+1+1+2+10+10=25 for our 10-bin analysis. Note that the total matter density is a sum of cold dark matter density and baryon density; the former feeds into \ccl; $\Omega_b, h_0^2$ are specified given our fiducial cosmology presented in \autoref{sec: map generation}.
\label{tab: priors}}
\end{table*}

Our likelihood is assumed to be Gaussian and is calculated using \lsslike\footnote{\url{https://github.com/LSSTDESC/LSSLike}}. To sample the posterior, we run \ttt{emcee}\footnote{\url{https://github.com/dfm/emcee}} \citep{emcee}, which implements an ensemble sampler for Markov chain Monte Carlo, with 100 walkers, 2000 steps for burn-in and 1000 steps for post-burn-in; all the walkers start from small perturbations to the true values.\footnote{When analyzing real data, we would start at random positions in the parameter space allowed by the priors or using perturbations to coarse estimates using e.g., a likelihood minimizer. Here, we proceed by slightly perturbing the truth (by 10\% of the prior width) for a faster analysis; this should not be detrimental here since this proves sufficient to trace out the likelihood near its optimum, and we perform only a relative analysis of the resulting constraints.} All our chains pass the Gelman-Rubin convergence criteria \citep{gelman-rubin}; a few also pass the Geweke one \citep{geweke}. We also visually inspect all chains to ensure stability. In a couple of cases, however, there are stuck walkers which we eliminate by implementing a clipping criterion across all cases\footnote{We achieve this by discarding walkers that are clear outliers: with paths with log probability whose distance from the median log probability is $\geq$ 10 times the median distance. This leads to only one case with 4/100 walkers discarded while four cases have 1/100 walkers discarded; the rest have no walkers discarded. Note that given the small number of walkers clipped (i.e., 1/100 for 4/20 cases and 4/100 for 1/20 cases), there is no practical impact aside from cleaner posterior plots.
}.

\section{Systematics: Cases\label{sec: systematics}}
In this section, we describe the models for the various systematics that we study in this work. To make clear the different working parts, we present the various cases in \autoref{tab: cases}. Essentially, for every case, we have an input \pz\ model, an input Milky Way (MW) dust model, a set of template contaminants, and the covariance used for the likelihood. While the observed galaxy density maps are generated with case-specific details (i.e., specific photometric redshift distribution, MW dust model and observational artifacts, as the output of \autoref{sec: observed maps}), we assume no knowledge of them beyond the baseline models when carrying out the likelihood analysis -- mimicking the real-world scenario of our imperfect knowledge.

\input{table2}
\subsection{Photometric Redshifts\label{sec: pz models}}
Here, we define the photometric redshift error distribution $\mathcal{P}(z_p; z)^{\mathrm{unnormed}}$ presented in \autoref{sec: pz-nz-model}, via \autoref{eq: joint pz general}, for the various cases that test various \pz\ systematics.
\subsubsection{Baseline\label{sec: pz baseline}}
Following Equation 13.9 in \citetalias{abell+2009}, we have
\eq{
	\mathcal{P}(z_p | z) \propto \begin{cases}
							\ \mathrm{exp}\bsqbr{ - \frac{(z_p - z - \delta_z)^2}{2 \sigma_z^2} } & z_p \geq 0 \\
							\ 0 & z_p < 0
						\end{cases}
	\label{eq: pz model}
}
where $z_p$ is the \pz\ (i.e., a \pz\ point statistic), $\delta_z$ is the \pz\ bias which we assume to be zero in the \baseline\ case, and $\sigma_z$ is \pz\ error. We follow \citetalias{desc-srd2018} and assume $\sigma_z \equiv \sigma_{z_0} (1+z)$ where $\sigma_{z_0}$ defines the \pz\ scatter amplitude for the target \pz\ calibration, with $\sigma_{z_0}=0.03$ as the baseline.
The probability of assigning a galaxy to a photometric bin $i$ with bin edges $\zpmin, \zpmax$ then is
\eq{
	\mathcal{P}(z_p \in [\zpmin, \zpmax) | z)
	= \frac{ \mathcal{I}(z_p \in [\zpmin, \zpmax) | z) }{ \mathcal{I}(0, \infty | z) }
	\label{eq: joint pz}
}
where
\eq{
	\mathcal{I}(z_p \in [a, b) | z)
	=  \frac{1}{\sqrt{2\pi}\sigma_z} \int_{a}^{b}
		dz_p' \ \mathrm{exp}\bsqbr{ - \frac{(z_p' - z - \delta_z)^2}{2 \sigma_z^2} }
	\label{eq: pz}
}
For the baseline case, we assume no spectroscopic calibration biases (neither additive nor multiplicative i.e., $\{\Delta_i$, $s_i\}$ = $\{0, 1\}$ in \autoref{eq: nz_i delta} and \autoref{eq: nz_i s}); this is in addition to details mentioned above: no \pz\ bias (i.e., $\delta_z = 0$) and baseline scatter amplitude (i.e., $\sigma_{z_0} = 0.03$). Note that these four parameters are changed depending on the systematics case, as discussed in the following subsections, while we employ the shift and stretch formalism in inference as described in \autoref{sec: estimation of params}; the former are in play in Column 3 in \autoref{tab: cases} and latter in Column 6. 

\begin{figure}[!htb]
	\centering
		\includegraphics[width=.5\linewidth]{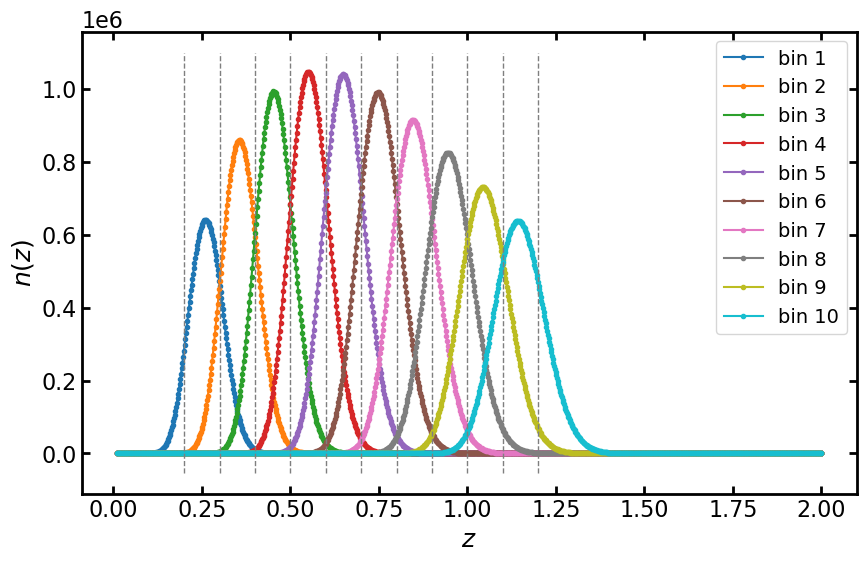}
	\caption{Number distribution of galaxies with the \baseline\ model for the \pz\ error distribution (no bias or outliers; scatter amplitude set to 0.03) and spectroscopic calibration biases (none) for the 10-bin \baseline\ case. The different colors represent different redshift bins, edges of which are shown as dotted vertical lines.}
	\label{fig: nz_i baseline}
\end{figure}

\autoref{fig: nz_i baseline} shows the galaxy distribution in our redshift bins resulting from using the baseline \pz\ distributions defined here.

\subsubsection{Additive Spectroscopic Calibration Bias\label{sec: pz shift}}
\begin{figure*}[!htb]
	\centering
	\begin{minipage}{0.49\linewidth}
		\includegraphics[width=\linewidth]{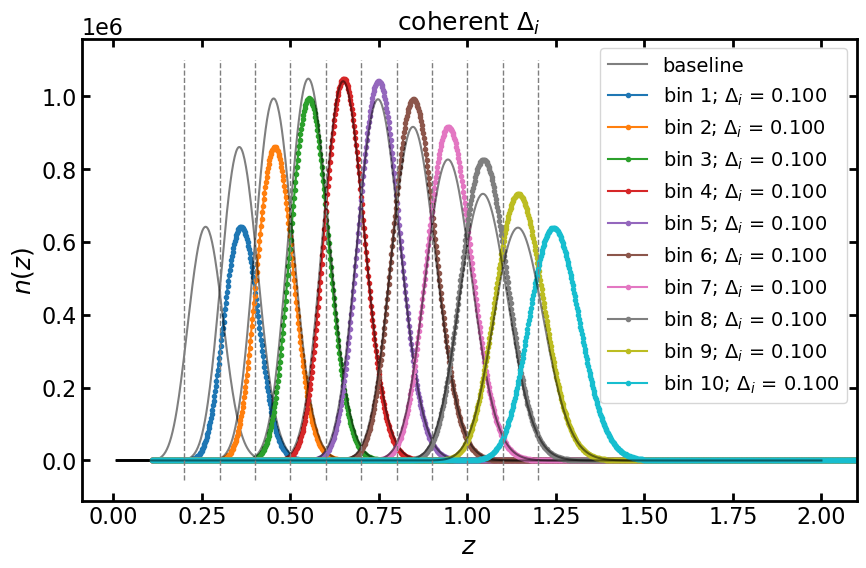}
	\end{minipage}\
	\begin{minipage}{0.49\linewidth}
		\includegraphics[width=\linewidth]{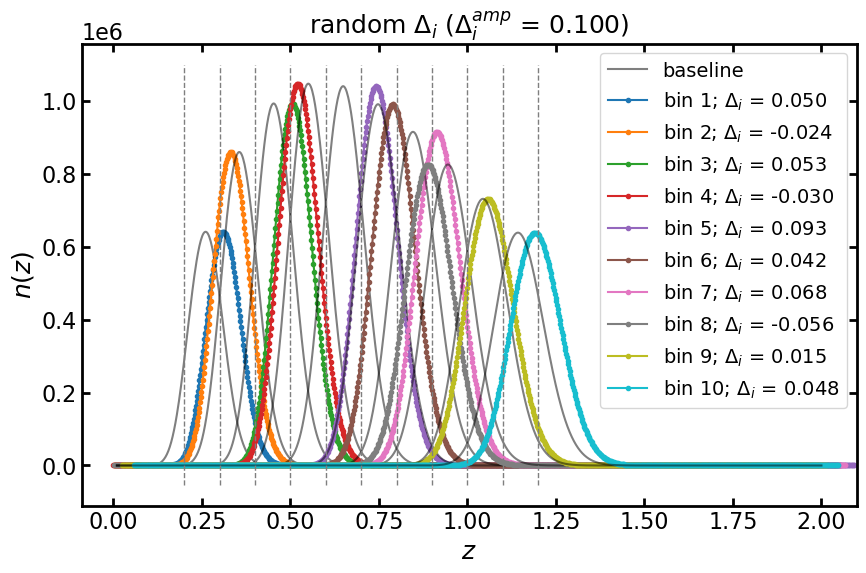}
	\end{minipage}\
	\caption{Number distribution of galaxies for the two additive \pz\ spectroscopic calibration bias cases, implemented via the shift parameter: coherent biases (left; with $\Delta_i=0.1$), and random ones (right; random $\Delta_i$ between -0.1, 0.1). Both panels show the baseline from \autoref{fig: nz_i baseline} in solid grey for comparison. Coherent bias with $\Delta_i=-0.1$ essentially shifts everything in the left panel to the left, instead of right.}
	\label{fig: nz_i pz-shift}
\end{figure*}

Here, we assume the same \pz\ error distribution as in \baseline, i.e., as described by \autoref{eq: pz model}, but add a non-zero spectroscopic calibration additive bias, $\Delta_i$. We implement coherent shifts, whereby all ten redshift bins get the same additive bias, as well as random shifts, where each bin gets its own bias though within the confines of a specified amplitude, $\Delta_i^{amp}$. \autoref{fig: nz_i pz-shift} shows the resulting galaxy number distributions for the two cases, with a maximum plausible amplitude of each ($\pm$ 0.1, the redshift bin width for the tomography considered here).

\subsubsection{Multiplicative Spectroscopic Calibration Bias\label{sec: pz stretch}}
\begin{figure}[!htb]
	\centering
	\begin{minipage}{0.49\linewidth}
		\includegraphics[width=\linewidth]{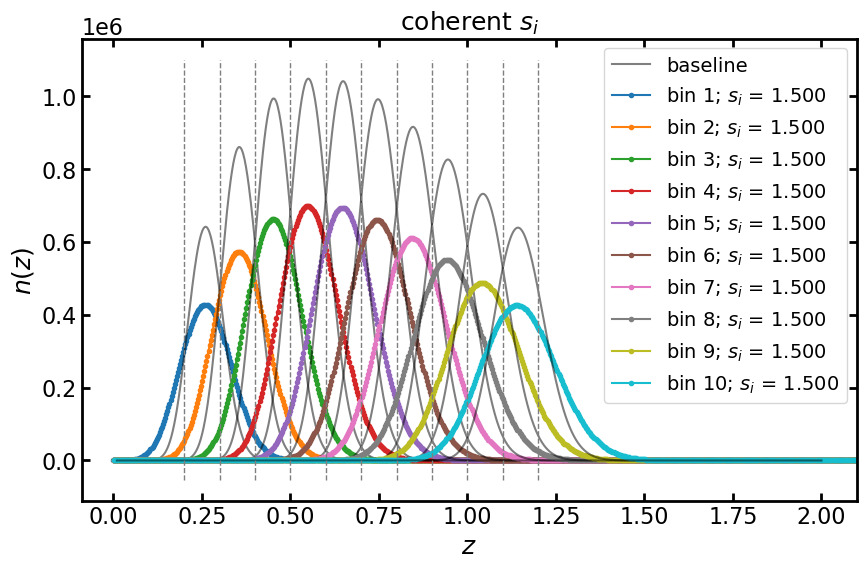}
	\end{minipage}\
	\begin{minipage}{0.49\linewidth}
		\includegraphics[width=\linewidth]{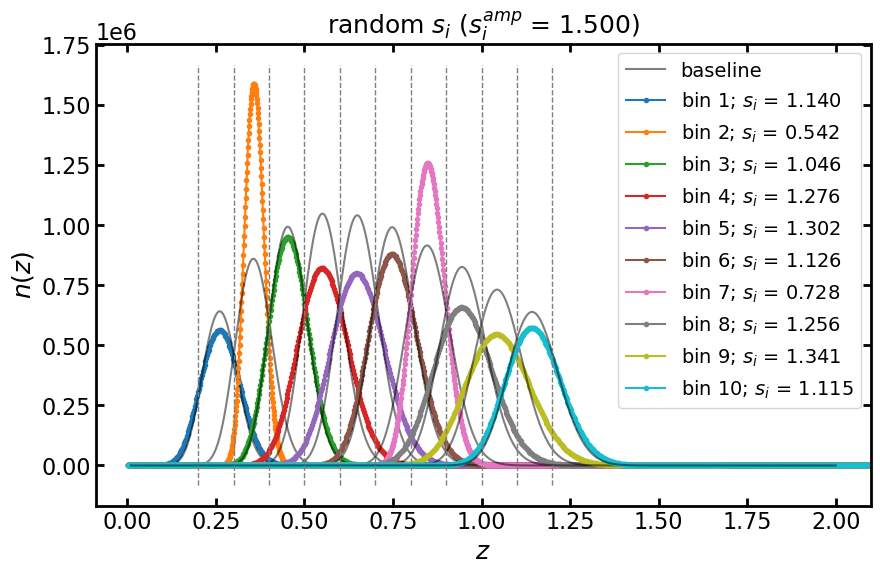}
	\end{minipage}\
		\caption{Number distribution of galaxies for the two multiplicative \pz\ spectroscopic calibration biases, implemented via the stretch parameter: coherent biases (left; with $s_i=1.5$), and random ones (right; random $s_i$ between 0.5, 1.5). Both panels show the baseline from \autoref{fig: nz_i baseline} in solid grey for comparison. We do not show coherent bias $<1$ but $s_i=0.75$ essentially makes everything in the left panel narrower, instead of wider.}

	\label{fig: nz_i pz-stretch}
\end{figure}

Here, we assume the same \pz\ error distribution as in \baseline, i.e., as described by \autoref{eq: pz model}, but add a non-zero spectroscopic calibration multiplicative bias, $s_i$. Similar to \autoref{sec: pz shift}, we implement coherent stretches, whereby all ten redshift bins get the same multiplicative bias, as well as random ones, where each bin gets its own stretch. \autoref{fig: nz_i pz-stretch} shows the resulting galaxy number distributions for the two cases, with a maximum plausible amplitude of each.

\subsubsection{Biased $\mathcal{P}(z_p|z)$\label{sec: pz bias}}
\begin{figure}[!htb]
	\centering
	\begin{minipage}{0.49\linewidth}
		\includegraphics[width=\linewidth]{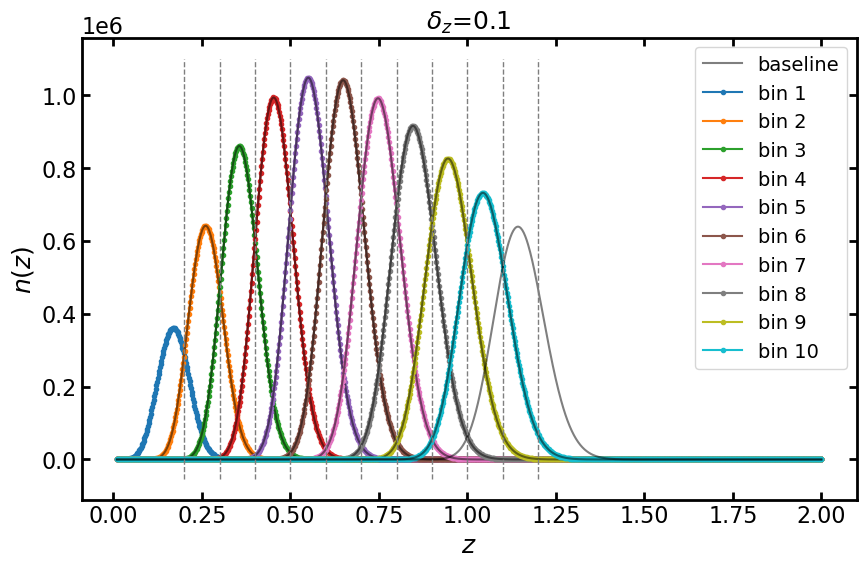}
	\end{minipage}\
	\begin{minipage}{0.49\linewidth}
		\includegraphics[width=\linewidth]{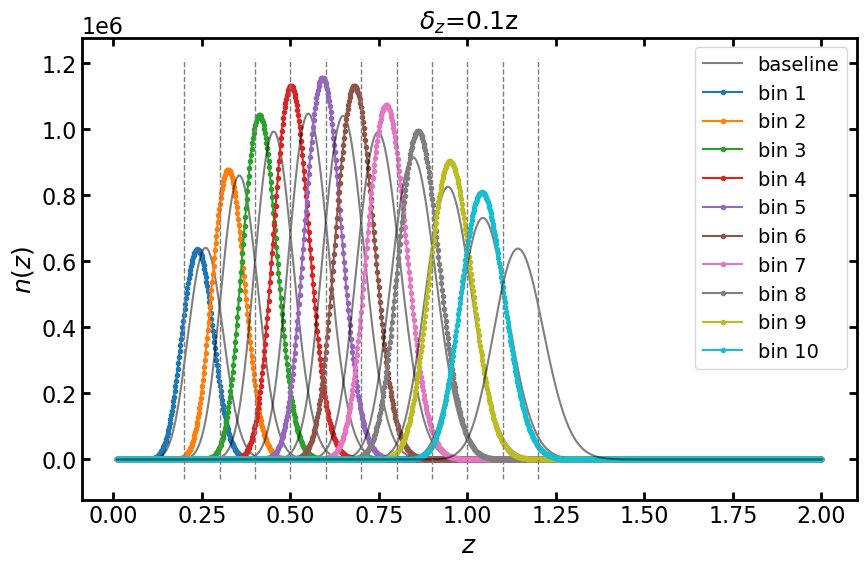}
	\end{minipage}\
	\caption{Number distribution of galaxies for the two \pz\ bias cases: constant bias (left; with  $\delta_z=0.1$), and linear bias (right; with  $\delta_z = 0.1z$). Both panels show the baseline from \autoref{fig: nz_i baseline} in solid grey for comparison.}
	\label{fig: nz_i pz-bias}
\end{figure}

Here, we assume the same \pz\ error distribution as in \baseline, i.e., as described by \autoref{eq: pz model}, but with two variants: 1) a constant bias implemented by using a specified value of $\delta_z$, and 2) a linear bias by using $\delta_z \propto z$. \autoref{fig: nz_i pz-bias} shows the resulting galaxy number distributions for the two cases for our 10-bin run, with a maximum plausible amplitude of each (0.1 and 0.1$z$, respectively, as our redshift bins are 0.1 wide).

\subsubsection{Large Photo-$z$ Scatter\label{sec: pz scatter}}
\begin{figure}[!htb]
	\begin{minipage}{0.49\linewidth}
		\includegraphics[width=\linewidth]{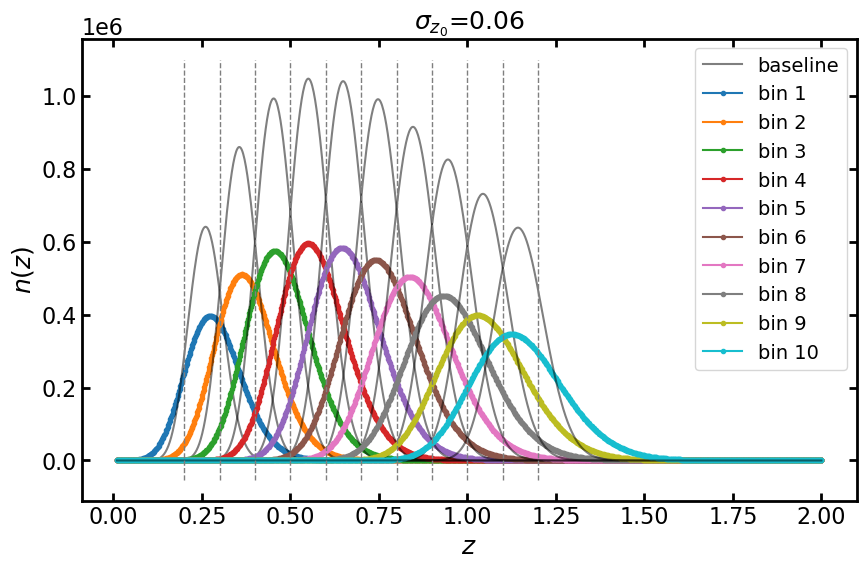}
		\caption{Number distribution of galaxies with \pz\ error distribution with a larger scatter amplitude than in \baseline. Baseline from \autoref{fig: nz_i baseline} in solid grey for comparison.}
		\label{fig: nz_i large-scatter}
	\end{minipage}\
	\begin{minipage}{0.49\linewidth}
		\includegraphics[width=\linewidth]{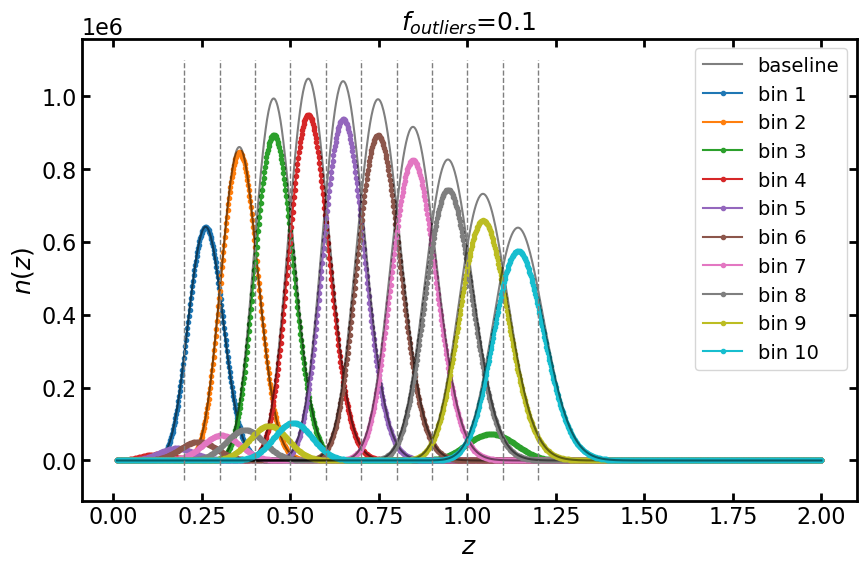}
		\caption{Number distribution of galaxies with \pz\ error distribution that has 10\% outliers, aliased across all redshift. Baseline from \autoref{fig: nz_i baseline} in solid grey for comparison.}
		\label{fig: nz_i pz outliers}
	\end{minipage}\
\end{figure}

Here, we assume the same \pz\ error distribution as in \ttt{baseline}, i.e., as described by \autoref{eq: pz model}, but with $\sigma_{z_0}$ larger than the baseline value. \autoref{fig: nz_i large-scatter} shows the resulting galaxy number distribution for this case, with a maximum plausible amplitude (2$\times$ the baseline value, 0.03).

\subsubsection{Photo-$z$ Outliers\label{sec: pz outliers}}
Photo-$z$s are prone to degeneracies between various spectral features. This happens when any two features can activate the same filter and we have:
\eq{
	\lambdaobs = \lambdaem{1}(1+z_1) = \lambdaem{2}(1+z_2)
}
\eq{
	\Rightarrow \frac{ 1+z_1}{ 1+z_2 } = \frac{ \lambdaem{1} }{ \lambdaem{2} }
	\label{eq: aliasing definition}
}
As an example relevant for our redshift range, consider the 4000\AA\ break and the Magnesium-II 2800\AA\ break where we can have aliasing between $z_1 \in [0.4, 0.5]$ and $z_2 \in [1.0, 1.14]$.

In order to model the impact of redshift aliasing, we modify \autoref{eq: pz model} as follows:
\eq{
	\mathcal{P}(z_p | z) \propto
	\begin{cases}
		\ f_\mathrm{outliers} \ \mathrm{exp}\bsqbr{ - \frac{(z_p^\mathrm{aliased} - z - \delta_z)^2}{2 \sigma_z^2} }
				& \begin{array}{@{}l@{}} z_p \geq 0,
				   z_{p}^\mathrm{min-aliased} \leq z_p \leq z_{p}^\mathrm{max-aliased}
				   \end{array}
				   \\
		\ (1-f_\mathrm{outliers}) \mathrm{exp}\bsqbr{ - \frac{(z_p - z - \delta_z)^2}{2 \sigma_z^2} }
		& z_p \geq 0,
		   z_p < z_{p}^\mathrm{min-aliased} \ \mathrm{or} \
		   z_p > z_{p}^\mathrm{max-aliased}
		\\
		\ 0 & z_p < 0
	\end{cases}
	\label{eq: pz model outliers}
}
where $(z_{p}^\mathrm{min-aliased}, z_{p}^\mathrm{max-aliased})$ marks the range that is prone to aliasing; $z_p^\mathrm{aliased}$ is given by \autoref{eq: aliasing definition}; and $f_\mathrm{outliers}$ is the outlier fraction. \autoref{fig: nz_i pz outliers} shows the resulting galaxy number distribution for a maximum plausible outlier fraction (10\%).

\subsection{Milky Way Dust\label{sec: mw models}}
\subsubsection{Baseline\label{sec: mw baseline}}
Our baseline MW dust model assumes the dust extinction map from \citet{sfd1998} (referred to as the SFD map) while we implement the dust law from \citet{cardelli+1989}\footnote{\url{https://github.com/lsst/sims_photUtils/blob/main/python/lsst/sims/photUtils/Sed.py\#L921-L979}}. Specifically, the dust law dictates that $A_\lambda / A_V = a(x) + b(x)/R_V$, where $A_\lambda$ is the wavelength-dependent extinction, and $x = 1 / \lambda$. Note that the reddening $E(B-V)$ is defined as $A_B - A_V$, while $A_V = R_V \times E(B-V)$. For \baseline, we assume $R_V = 3.1$.

\begin{figure}[!htb]
	\centering
		\includegraphics[width=0.5\linewidth, trim={5 5 5 5}, clip=false]{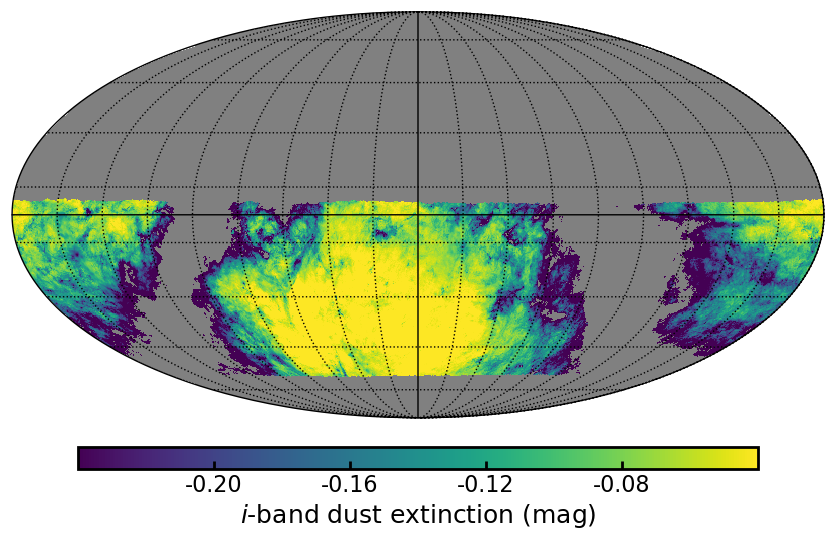}
	\caption{Mollweide projection of the $i$-band dust extinction map for the \baseline\ case, following the baseline Milky Way dust model (\autoref{sec: mw baseline}). We focus on the survey footprint of interest (i.e., that limited by depth and extinction, and with coverage in all six bands; explained in \autoref{sec: observed maps}); the rest is masked (appearing as grey).}
	\label{fig: baseline dust extinction}
\end{figure}

We incorporate the dust systematics by considering them alongside $\delta_\mathrm{OS}$ (as explained in \autoref{sec: observed maps}). \autoref{fig: baseline dust extinction} shows the dust extinction in the $i$-band for \baseline.
\subsubsection{Resolution Mismatch\label{sec: mw res-mismatch}}
Here we consider the impact of using a dust map whose resolution is lower than that of the data, which is a realistic concern given that the SFD angular resolution is 7 arcminutes \citealt{sfd1998} compared to the arcsecond resolution of LSST data. In order to simulate this, we keep \nside\ 1024 maps (with equal-area pixels of 11.8 arcmin$^2$ and consequent effective square-pixel side length of 3.44 arcmin) for the generation of the observed maps but use an \nside\ 64 dust map (with equal-area pixels of 3021.5 arcmin$^2$ and consequent effective square-pixel side length of 54.97 arcmin), interpolated to \nside\ 1024 using \healpy\ routine \ttt{ud\_grade}, when constructing the template to be passed to \nmt. We expect that when correcting with the lower-resolution dust map, we will see excess power on small scales (i.e., large $\ell$s); the incorrect small-scale dust correction will lead to extra small-scale apparent galaxy clustering.

\begin{figure}[!htb]
	\centering
    \hspace*{-0.6em}
	\begin{minipage}{0.43\linewidth}
		\includegraphics[width=\linewidth]{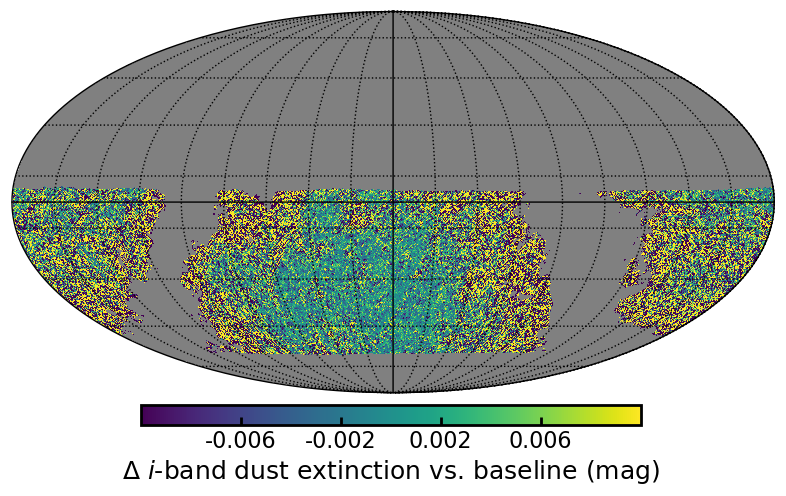}
	\end{minipage}\
    \hspace*{-0.5em}
	\begin{minipage}{0.58\linewidth}
		\includegraphics[width=\linewidth]{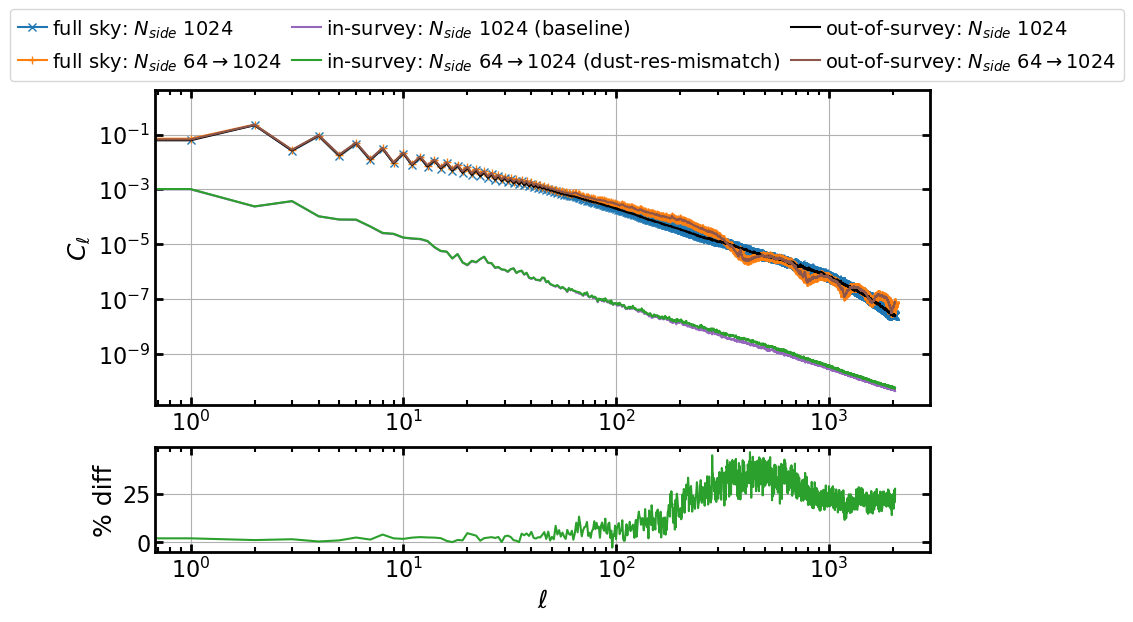}
	\end{minipage}
    \hspace*{-0.5em}
	\caption{
	\textit{Left}: Skymap showing the difference in the dust extinction when comparing \baseline\ vs. the resolution mismatch case described in \autoref{sec: mw res-mismatch}.
	\textit{Right}: Power spectrum of the full-sky reddening map\protect\footnotemark\ at high resolution vs. that constructed from low resolution (blue vs. orange), for in-survey footprint (purple vs. green), and out-of-survey footprint (black vs. brown); the lower panel shows the \% relative difference from \baseline. We see that the interpolation using the lower resolution map is not perfect at reproducing the true power on small scales but our survey footprint is largely immune from the excess power.
	}
	\label{fig: res-mismatch}
\end{figure}

The left panel in \autoref{fig: res-mismatch} shows the difference in dust extinction compared to \baseline\ where we see that the resolution change introduces some structure. In order to understand the scale of the induced structure, we plot the angular power spectra, shown in the right panel in \autoref{fig: res-mismatch}, where we see that our in-survey footprint (defined in \autoref{sec: del-os}) is largely immune from the impacts of the interpolation. Therefore, we expect that the excess structure introduced due to a lower dust map resolution will not have dire impacts on our posteriors.

\footnotetext{We use the reddening maps instead of the dust extinction maps since MAF, the simulation analysis framework, does not support full-sky dust extinction maps with non-full-sky observations scheduled (as for LSST) and we want to see the effects of the resolution on full-sky vs not (since only then we see the low-resolution playing a role).}


\subsubsection{Incorrect Dust Law\label{sec: mw rv change}}
\begin{figure}[!htb]
	\centering
	\begin{minipage}{0.49\linewidth}
		\includegraphics[width=\linewidth]{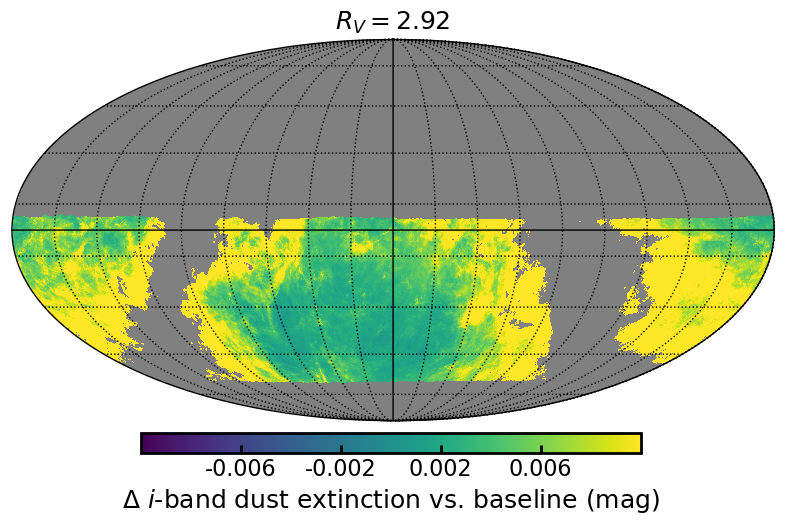}
	\end{minipage}\
	\begin{minipage}{0.5\linewidth}
		\includegraphics[width=\linewidth]{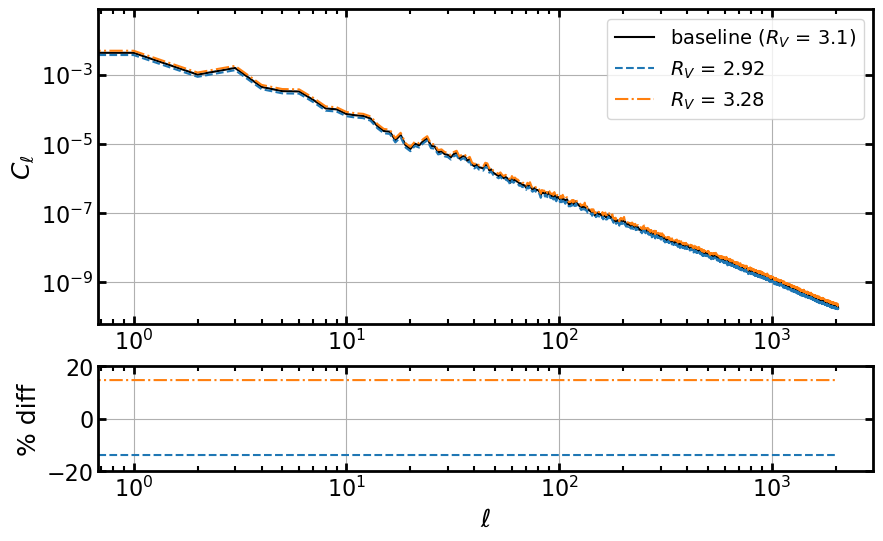}
	\end{minipage}\
	\caption{
    \textit{Left}: Skymap showing the difference in the dust extinction when changing the dust law to $R_V$ = 2.92 vs. \baseline\ ($R_V=3.1$); the skymap with $R_V=3.28$ is essentially mirrored, with $R_V$ = 3.28 leading to more dust extinction compared to $R_V$ = 2.92. We see that the changes are sub-percent level in magnitude.
	\textit{Right}: Power spectrum of the dust extinction maps with differing dust laws: $R_V=3.1$ (baseline; black), $2.92$ (blue), and $3.28$ (orange); the lower panel shows the \% relative difference from \baseline. We see that the impacts are small but present.
    }
	\label{fig: rv change}
\end{figure}

As a simple deviation from the baseline dust law, we generate observed density maps with a different value for $R_V$ than in \baseline. We try both a positive and negative excursion in the $R_V$ value within its uncertainty range of 0.18 \citep{schlafly+2016}. \autoref{fig: rv change} shows the change in dust extinction between that resulting from $R_V = 3.1$ (baseline) vs. $R_V = 2.92$ in the left panel, and the angular power spectra of the dust extinction maps in the right one. We see that the changes are small, leading us to expect minimal impacts on our posteriors.

\subsection{Contaminant Templates\label{sec: ct}}
In order to account for various contaminants in our data, we pass templates to \nmt\ to deproject, alongside the density maps for the specific redshift bins. The goal is to test the impact of individual contaminant templates.

As our \baseline, we pass three contaminant templates: 1) the dust-corrected $i$-band coadded 5$\sigma$ depth accounting for dust extinction, 2) $i$-band seeing, and 3) the artificial LSS map that is induced due to observational systematics aside from MW dust, denoted as $\delta_\mathrm{OS}^\mathrm{no\ dust}$.

\begin{figure}[!htb]
	\centering
	\includegraphics[width=\linewidth, trim={5 5 5 5}, clip=false]{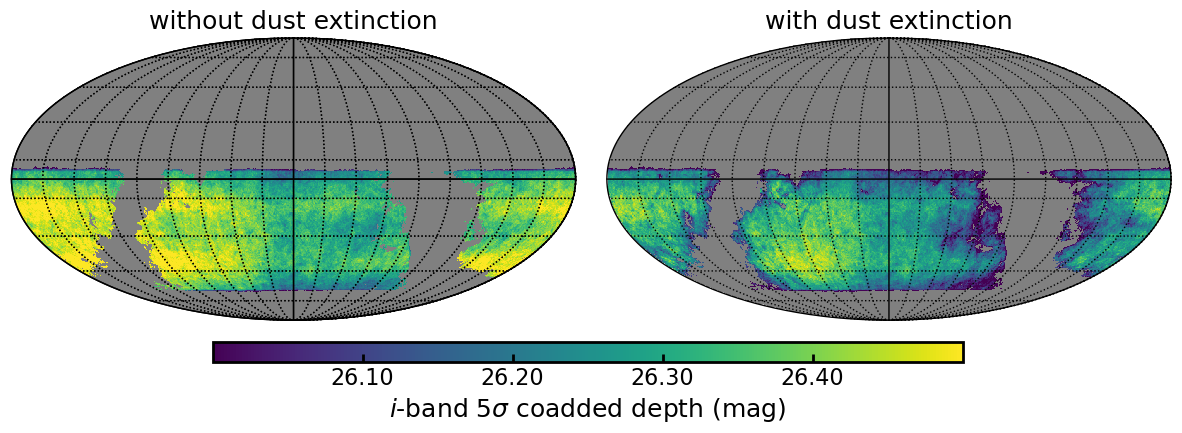}
	\caption{Skymaps showing the (Year 10) $i$-band coadded 5$\sigma$ depth map without accounting for dust extinction (left) and with it (right); structure in the left panel arises from factors like the observing strategy.}
	\label{fig: coaddm5 with/without dust}
\end{figure}

The first map is calculated using maps of coadded depth and the dust extinction; \autoref{fig: coaddm5 with/without dust} shows the \baseline\ coadded depth maps before and after accounting for dust extinction (with \baseline\ dust extinction shown in \autoref{fig: baseline dust extinction}).

\begin{figure}[!htb]
	\centering
	\begin{minipage}{0.49\linewidth}
		\includegraphics[width=\linewidth, trim={5 5 5 5}, clip=false]{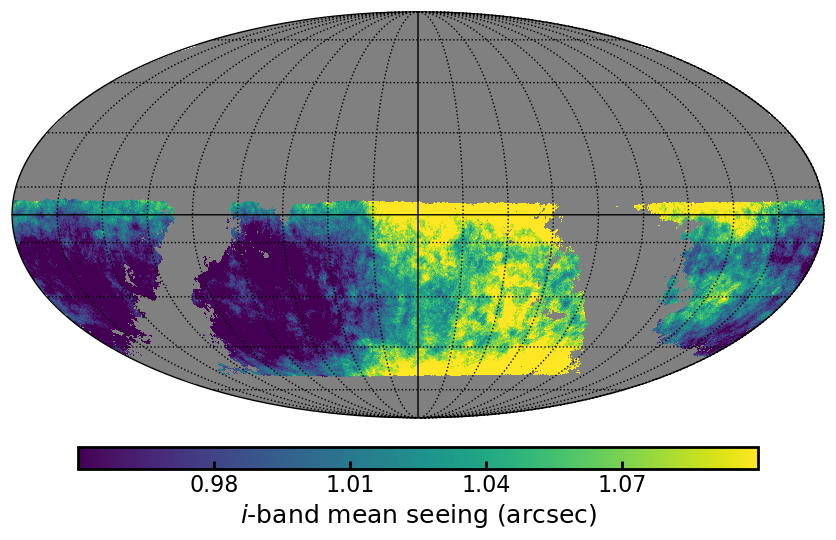}
	\caption{Skymap showing the mean seeing in $i$-band.}
	\label{fig: mean seeing}
	\end{minipage}\
	\hspace*{0em}
	\begin{minipage}{0.49\linewidth}
		\includegraphics[width=\linewidth, trim={5 5 5 5}, clip=false]{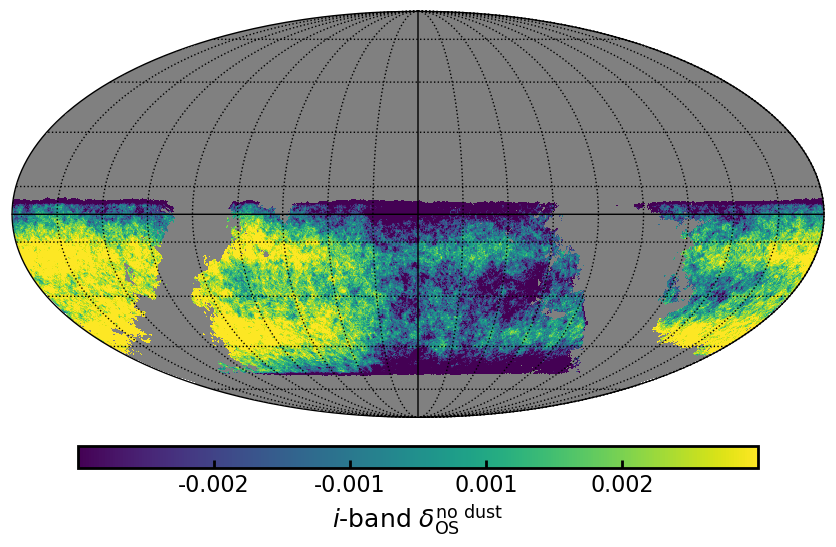}
		\caption{Skymap showing fluctuations due to observational systematics except dust.}
	\label{fig: del-os without dust}
	\end{minipage}\
\end{figure}

The seeing is based on the effective full-width at half maximum of the Point Spread Function (PSF)\footnote{See more at \url{https://rubin-scheduler.lsst.io/fbs-output-schema.html}}. \autoref{fig: mean seeing} shows the mean seeing in each \healpix\ pixel.

Finally, the artificial LSS induced due to the observational systematics aside from MW dust is calculated as discussed in \autoref{sec: del-os}; the quantity is shown in \autoref{fig: del-os without dust}.

For our test cases, we pass only one template to be deprojected, giving us three cases each with just one contaminant template: 1) $i$-band coadded 5$\sigma$ accounting for dust extinction, 2) seeing only, and 3) $\delta_\mathrm{OS}^\mathrm{no\ dust}$ only. We also add a case with passing just the dust extinction map, alongside a case with no templates (and, therefore, no deprojection).

\subsection{Covariance Matrix\label{sec: cov}}
\begin{figure}[!htb]
	\centering
	\includegraphics[width=\linewidth, trim={5 5 5 5}, clip=true]{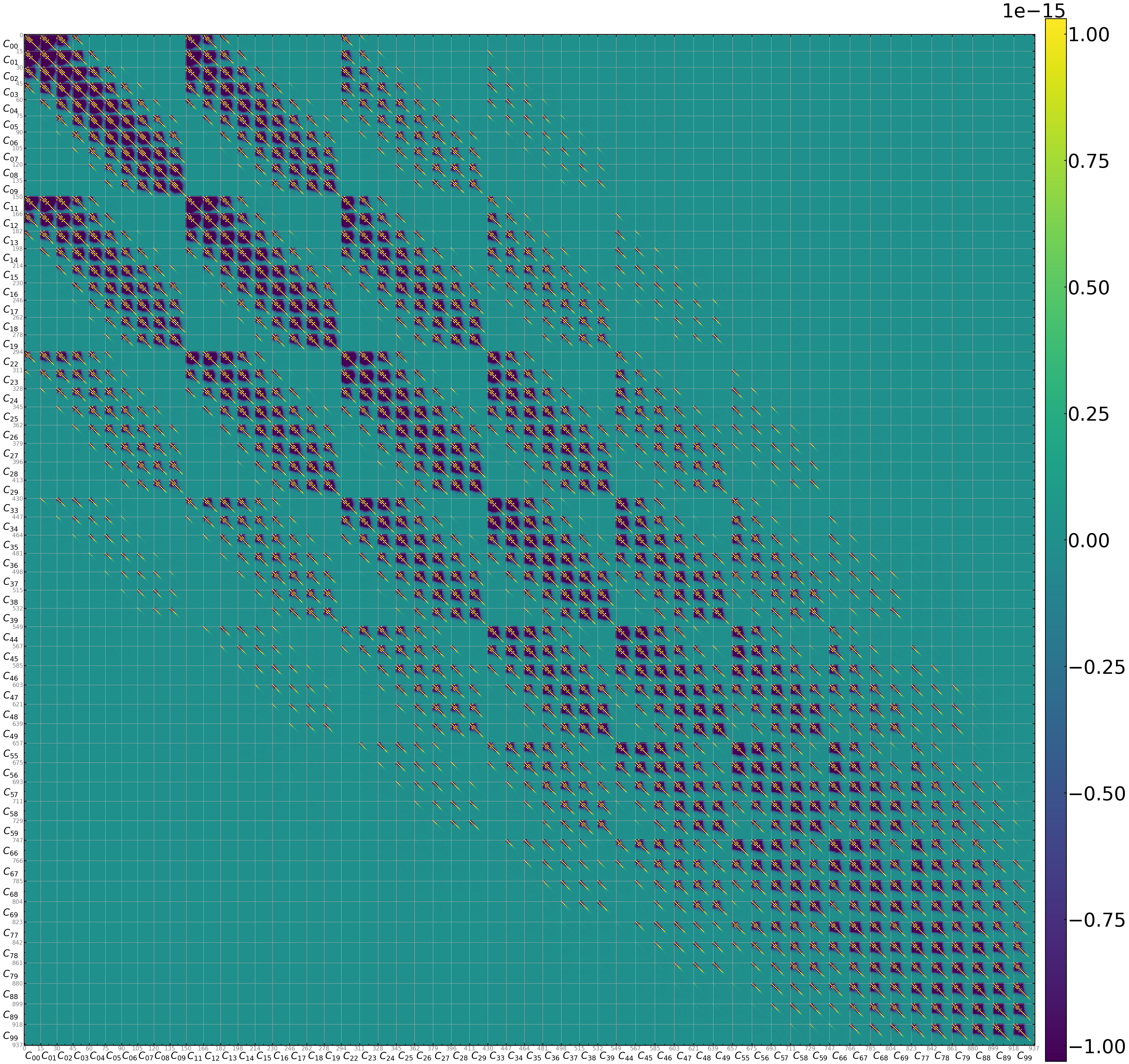}
	\caption{Full covariance matrix, outputted from \nmt.  }
	\label{fig: cov}
\end{figure}

For the baseline case, we use the full Gaussian covariance matrix outputted from \nmt. As a test case, we consider only the diagonal of this covariance matrix when running our likelihood analysis; the full covariance matrix is shown in \autoref{fig: cov}.

\FloatBarrier
\section{Results\label{sec: results}}
\begin{figure}[!htb]
	\centering
		\includegraphics[width=\linewidth, trim={5 5 5 5}, clip=true]{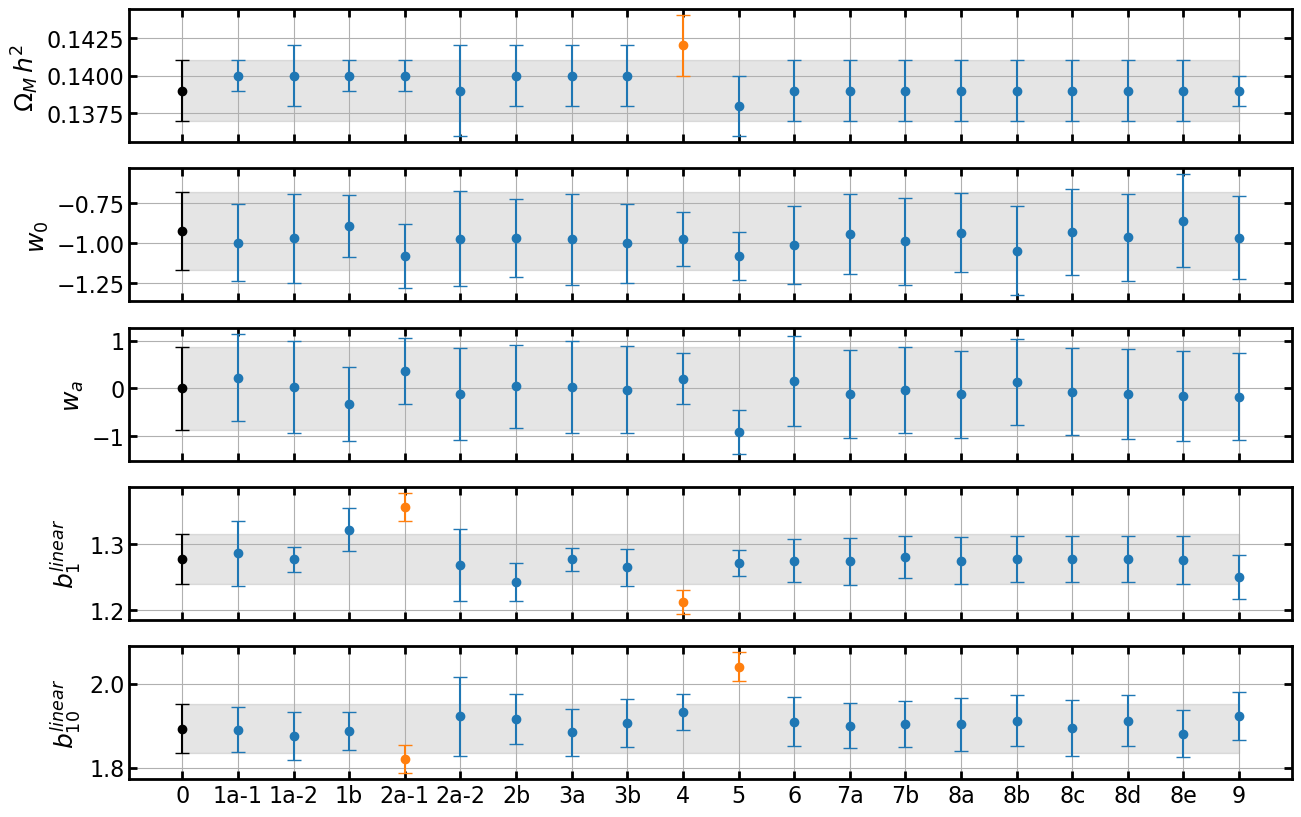}
	\caption{
    Best fit values for cosmology parameters for all cases considered here; see \autoref{tab: cases} or \autoref{tab: results1} for identifying case numbers with case details. Here, in each subplot, the grey band marks the uncertainty on \baseline\ best fits, shown as black points; orange points identify best fit values which are $>1\sigma$ away from baseline, as defined by \autoref{eq: bias def}; blue color is used for all the rest. \autoref{tab: results1} tabulates the data plotted here, alongside including the various statistics used to probe the fits and comparisons with \baseline. Analog figures for the rest of the parameters are included in \autoref{sec: bestfits}.
    }
	\label{fig: bestfit cosmoparams}
\end{figure}

While we discuss individual cases in the subsections below, we summarize our findings in \autoref{tab: results1} tabulating the best fits for some of the parameters  alongside various statistics to test the best fits and probe at the deviation between \baseline\ and any given case. \autoref{fig: bestfit cosmoparams} plots out the best fits from the table, while \autoref{fig: bestfit delzs} and \autoref{fig: bestfit szs} present the analogs for all the remaining parameters. Overall, we find mainly the \pz\ systematics leading to a bias in best fit values compared to \baseline, where bias is defined as 
\eq{
\frac{
    { \left| \ \mathrm{best\ fit}_\mathrm{case} - \mathrm{best\ fit}_\mathrm{baseline} \ \right| }
    }{
    \sqrt{\sigma_\mathrm{case}^2 + \sigma_\mathrm{baseline}^2}
    }
\label{eq: bias def}
}
where we take $\sigma$ to be the maximum between the upper and the lower best fit error\footnote{In the case where the truth is different for \baseline\ vs. a given case, we shift the best fit of the latter by the difference in truths. For example, for the case with additive spectroscopic bias of 0.1, the true value for $\Delta_i$ is 0.1, not 0 as for \baseline; in this case, we shift best fit$_\mathrm{case}$ by 0.1 before calculating the bias.}.

\subsection{Baseline\label{sec: results: baseline}}
\begin{figure}[!htb]
	\centering
		\includegraphics[width=\linewidth, trim={5 5 5 5}, clip=true]{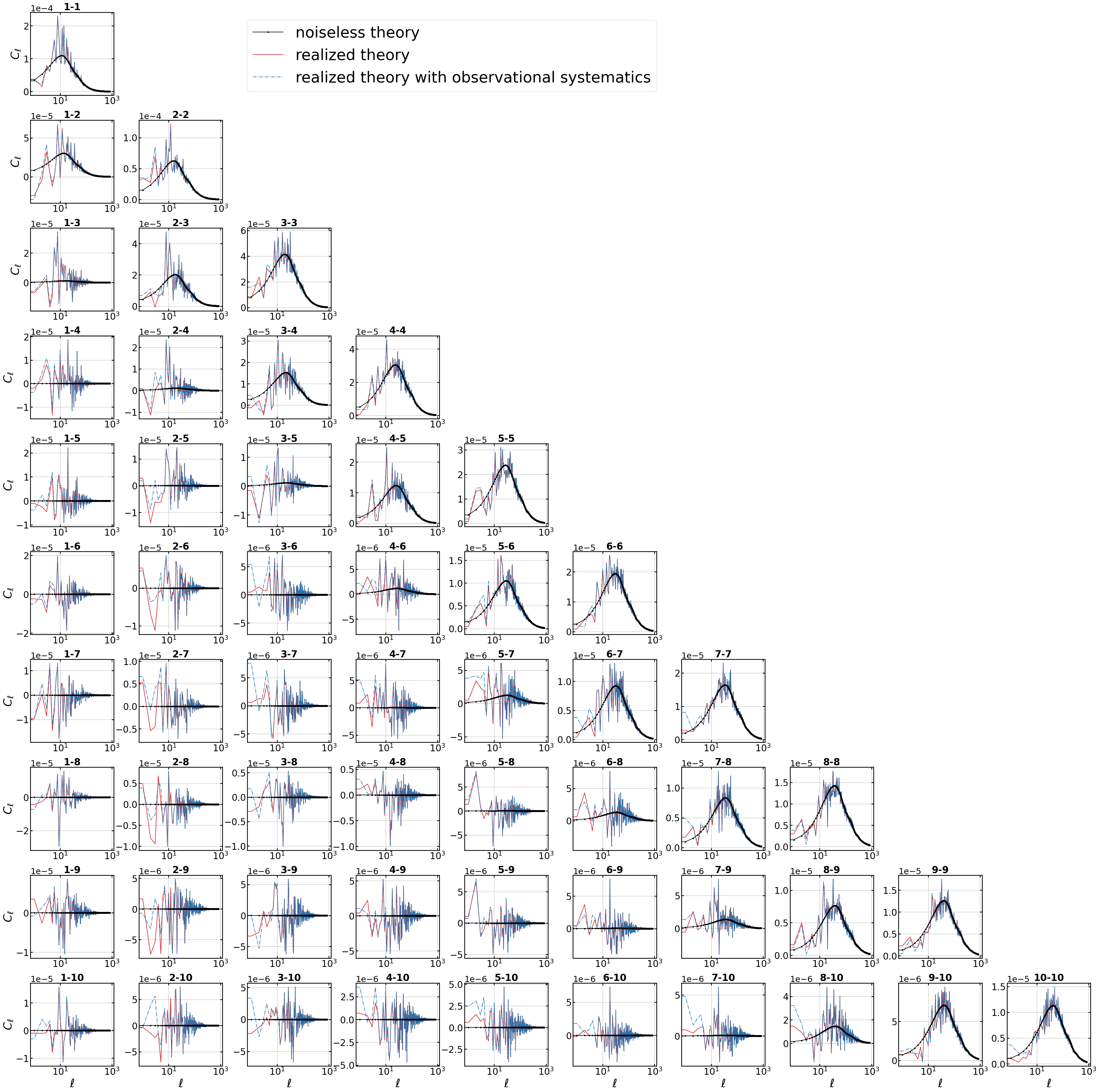}
	\caption{Simulated power spectra at various stages for the 10-bin case before running \nmt: pure (noiseless) theory from \ccl\ (black), spectra from realized theory maps (red), and those with added observational systematics (including dust and observing strategy; blue). The subplot titles show the redshift bins; ($C_\ell$) $i$-$j$ being the correlation between the density map for redshift bin $i$ and redshift bin $j$. Note that the $y$-range is different for each subplot.}
	\label{fig: 10bin interm spectra}
\end{figure}

Here, we show some details of the \baseline\ case. Specifically, \autoref{fig: 10bin interm spectra} shows the angular power spectra for the various stages leading to the simulated observed density maps. We see that the realized theory follows the input (noiseless) theory but has noise while the addition of observational systematics does not have substantial impacts; the realized spectra are based on density maps produced following the details in \autoref{sec: map generation} and \autoref{sec: observed maps}.

\begin{figure}[!htb]
	\centering
	\includegraphics[width=\linewidth, trim={5 5 5 5}, clip=true]{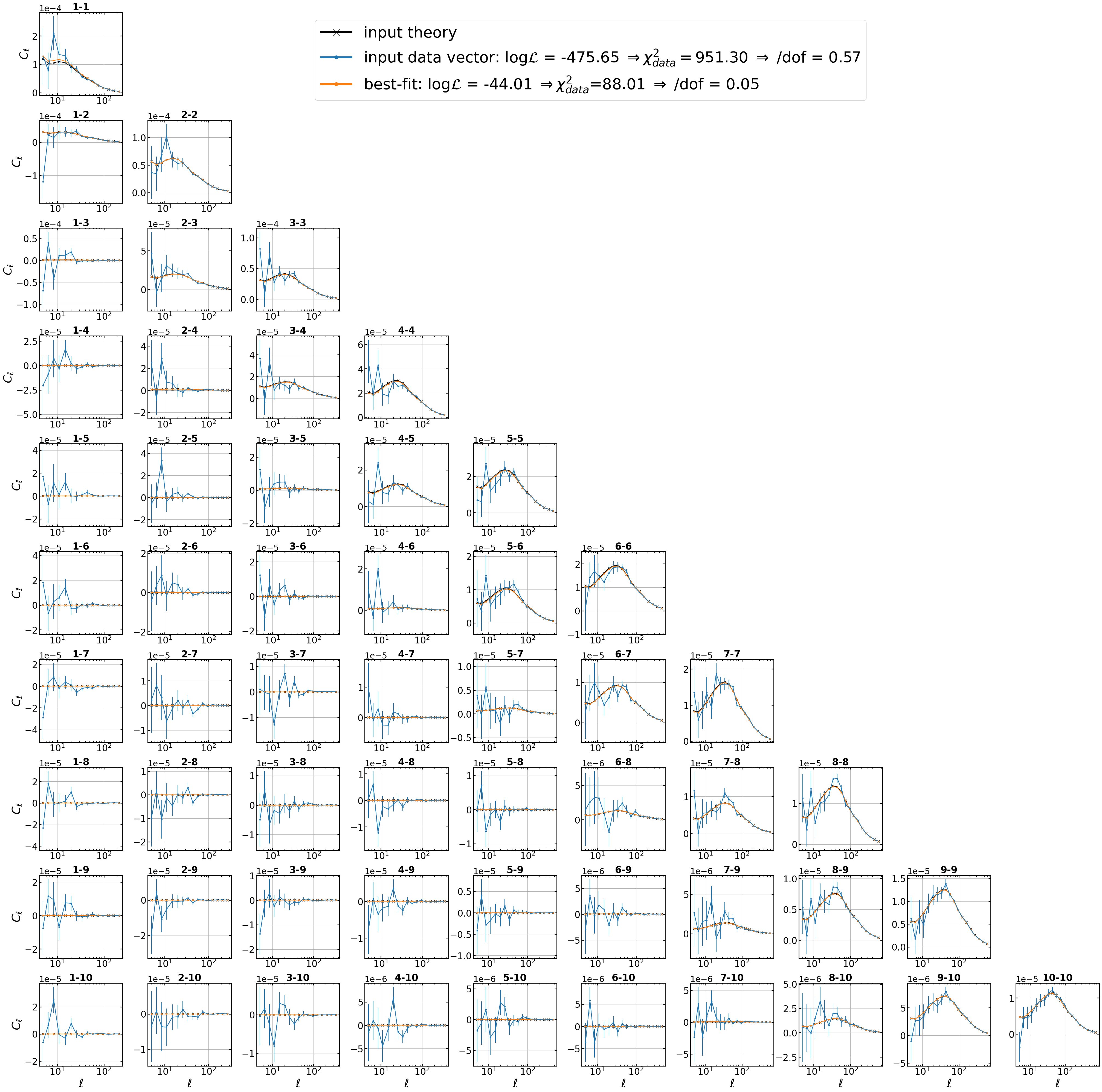}
	\caption{Power spectra for the 10-bin case at various stages related to inference: input (noiseless) theory (using \ccl; black), data vector (output from \nmt; blue), and those produced by \ccl\ using the best fit parameters from the inference analysis (orange). Both theory and best fit spectra are windowed and \lmax-clipped for a 1-1 comparison with the data vector. The legend shows \chitwodata\ for the fit of the respective vector; this is effectively the statistic used in the inference (=$-$2\logL). We see that best fit parameters (orange) lead to a smaller \chitwodata, implying a better fit than the data vector (blue). Note that we have a total of 1698 data points (i.e., the number of multipoles across all bins and spectra) which leads to 1672 degrees of freedom with 25 parameters being fit. Subplot titles are the same as in \autoref{fig: 10bin interm spectra}; $y$-axes limits are different in each subplot.}
	\label{fig: 10bin final spectra}
\end{figure}

As explained in \autoref{sec: estimation of cls}-\autoref{sec: estimation of params}, the observed density maps are passed to \nmt, which outputs decontaminated spectra (with mode coupling and windowing) and a Gaussian covariance matrix. These are then used in the likelihood analysis. \autoref{fig: 10bin final spectra} shows the theory spectra, the \nmt\ output spectra, as well as the spectra using the best fit params from the likelihood analysis. We see that the best fit parameters lead to spectra comparable to input theory (i.e., the truth); \chitwodata (=$-2$log$\mathcal{L}$) is better for the spectra from the best fit parameters as opposed to those outputted from \nmt.

\begin{figure}[!htb]
	\centering
		\includegraphics[width=\linewidth, trim={5 5 5 5}, clip=false]{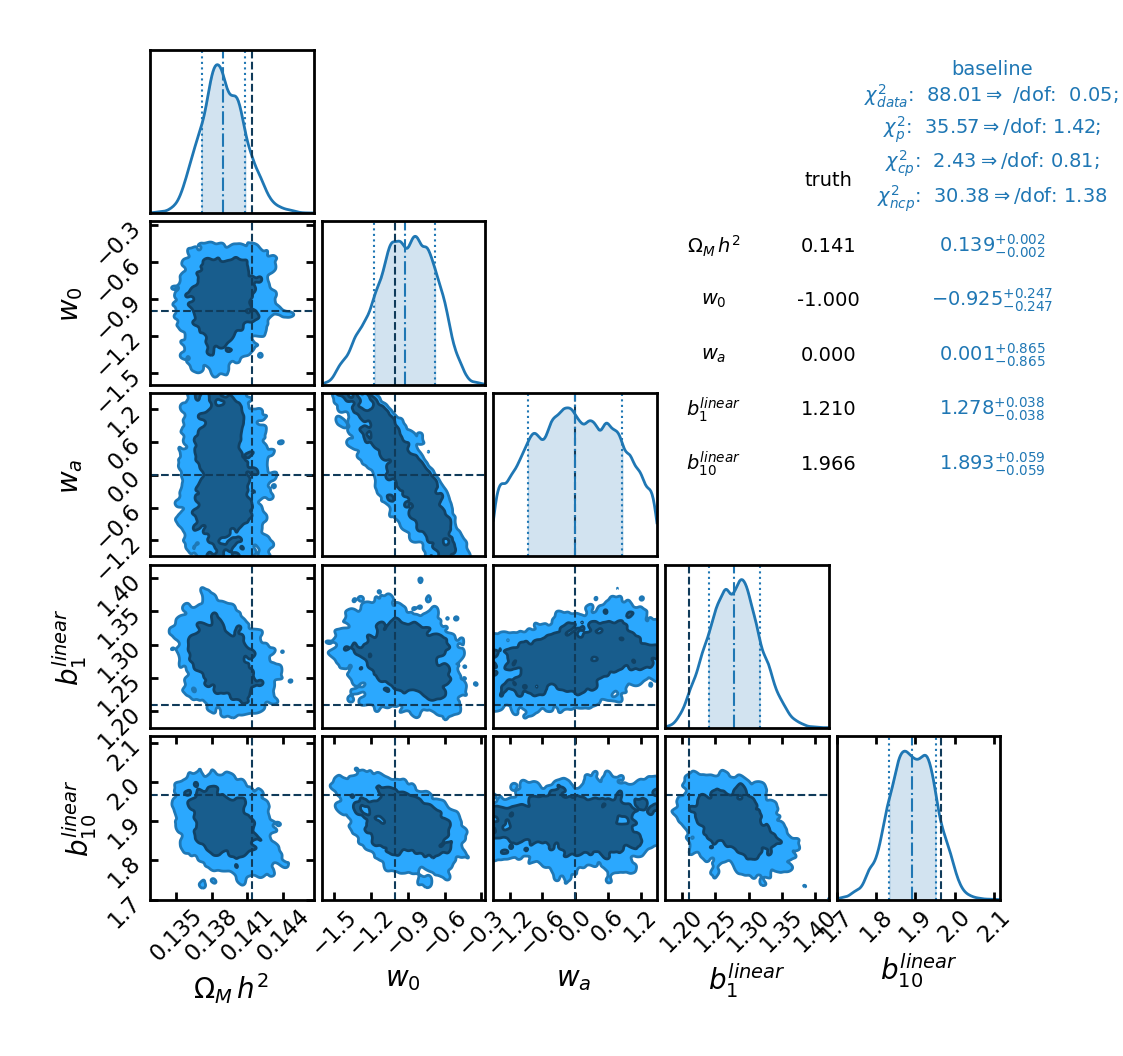}
	\caption{Baseline posteriors for 5 (out of 25) parameters fit for in our inference analysis; see \autoref{fig: posteriors baseline full} for all parameters. In the legend, we show the truth values for the parameters, and the best fit values (using the mean of the samples). We also include various statistics introduced in \autoref{tab: results1}. The black dashed line in the subplots denotes the truth; the dashed and dotted blue lines in the diagonal subplots indicate the best fit and best fit errors, respectively. 
    While some of the best fit parameter values (in the legend table) are more than 1$\sigma$ away from the truth, the deviation is never more than 2$\sigma$; the degree of deviation is even less pronounced in the 2D plots. While the best fits are not perfect, they establish the baseline against which we compare the rest of the cases. See the caption of \autoref{tab: results1} for details of the various \chitwo\ statistics in the legend.
    }
	\label{fig: posteriors baseline}
\end{figure}

Finally, \autoref{fig: posteriors baseline} shows a subset of posteriors from our likelihood analysis using the baseline models where we see that the best fit parameters are within 1-2$\sigma$ of the truth, establishing the baseline against which we compare the rest of the cases; see \autoref{fig: posteriors baseline full} for all the parameters. Note that we also test the impact of the randomness of the theory realization on the baseline posteriors, discussed in more detail in \autoref{sec: theory-seeds}. Using 10 different random seeds, we see deviations in best fit up to 2$\sigma$ (when compared to truth) -- but this should not impact our conclusions since we are fixing the seed for all analysis runs to be the same and only doing a comparative analysis (against \baseline).

\begin{landscape}
\input{table3}
\end{landscape}

\FloatBarrier
\subsection{\Pzs}
As described in \autoref{sec: pz shift}, we implement two cases of spectroscopic calibration additive bias: coherent ones and random ones; \autoref{fig: posteriors pz shift} shows the posteriors for these. We find that our $\chi^2$ statistics (\chitwodata, \chitwop, \chitwoncp; not \chitwocp) are drastically worse\footnote{Note that we intentionally choose to not be very quantitative with the comparisons of these statistics, given that their meaning is not absolute in the comparative framework of this work. Roughly speaking, however, ``drastically worse" covers multiple orders of magnitude differences, ``worse" encompasses up to a couple orders of magnitude differences, while ``not too affected" means less than order of magnitude difference. Note that in a real-data analysis, a reduced \chitwodata\ not close to 0 would indicate a problematic fit and would entail an investigation.} only with random additive biases, while coherent biases affect the statistics less drastically, implying that these biases are handled quite well by the \pz\ parameters. However, it is interesting to note that the shift parameters are not recovered perfectly for any of these cases (as most are $>1\sigma$ away from \baseline\ best fit).

As for the spectroscopic calibration multiplicative biases, for which we also try coherent and random cases, the posteriors are shown in \autoref{fig: posteriors pz stretch}. We find that the stretch parameters are never recovered perfectly, and in no configuration is the \chitwodata\ not adversely impacted while \chitwop\ only worsens for the first coherent case. The rest of the parameters are recovered reasonably however, except for the first coherent case, which leads to $>1\sigma$ biases in the bias parameters and one of the shift parameters.

As for the \pz\ bias, as described in \autoref{sec: pz bias}, we implement two cases. \autoref{fig: posteriors pz bias} shows the posteriors for the case with the constant \pz\ bias as well as the linear one. We find that our best fit parameters are mostly comparable with \baseline; the \chitwo\ statistics suffer but not drastically.

For larger \pz\ scatter amplitude, we find adverse impacts on quite a few of the best fit parameters, including cosmology ones; \chitwo\ statistics are not too adversely affected though, indicating that this systematic has the potential to bias our analysis without a clear indication of a problem with the test statistics; the posteriors for this case are shown in \autoref{fig: posteriors pz large-scatter}.

Finally, for \pz\ outliers, we test the impact with 10\% outliers; \autoref{fig: posteriors pz outliers} shows the posteriors. We find that quite a few of best fit values are biased, although not the core cosmology ones, against the \baseline, and the \chitwo\ statistics worsen, though still not as dramatically as in the calibration bias cases.

\subsection{MW Dust}
As discussed in \autoref{sec: mw res-mismatch}, we test the impacts of resolution mismatch between the dust extinction map used to produce our data vector vs. what is used in the deprojection. We find that the mismatch makes a negligible difference; \autoref{fig: posteriors mw res-mismatch} shows the posteriors for these cases.

Similar is the case for the incorrect dust law, for which the posteriors are shown in \autoref{fig: posteriors mw rv}. The mismatch between the true and assumed dust model does not show any significant differences.

\subsection{Contaminant Templates}
\autoref{fig: posteriors cts} shows the posteriors for the case where we pass no templates or one of the four described in \autoref{sec: ct}, instead of passing three out of the four in \baseline. We find that the results are only marginally different than \baseline, an interesting finding, which, while reassuring in terms of the impacts of observational systematics in the analysis framework here, may speak to the robustness of the survey details (Year 10 LSST data; survey footprint with strict cuts) -- but requires more detailed simulations to trust in the face of real object catalogs; we discuss this further in \autoref{sec: conclude}.

\subsection{Covariance Matrix}
\autoref{fig: posteriors cov diag} shows the posteriors for the case where we use only the diagonal of the full covariance matrix in our likelihood analysis, as described in \autoref{sec: cov}. We see that while the posteriors look slightly different, our test diagnostics do not show any significant changes, implying that most of the constraining power is coming from the diagonal.

\section{Conclusions \& Discussion\label{sec: conclude}}
In this work, we performed an end-to-end analysis of systematics that affect LSS galaxy clustering analysis. We used simulated, 10-year LSST galaxy density maps, with a focus on identifying whether the maximum plausible level of each systematic causes a measurable bias in our posteriors.  

Within the framework we used, whereby we decouple imaging systematics from others, we find that the most pressing systematics are those due to \pz s. In particular, we find that coherent additive spectroscopic calibration biases are not as problematic as random ones, even though the shift parameter (which captures the additive bias in our model) struggles even with coherent biases. The situation is quite different with multiplicative biases, which adversely impact the results regardless of the configuration. In both of these cases, when the fit for cosmological parameters is not good, it is reflected in \chitwodata, the \chitwo\ statistic of the data versus the best fit model, while \chitwocp, capturing the cosmology parameters covariance, doesn't worsen. In contrast, higher \pz\ scatter leads to biases in best fit without necessarily a larger \chitwodata, making this a more notorious systematic. \Pz\ bias and \pz\ outliers also have adverse impacts, although not as drastic ones; this is reassuring since we tested extreme cases of these systematics.

For other systematics, we find that we are robust to MW related systematics as well as observational systematics (traditionally mitigated via template deprojection). Specifically, using a lower-resolution dust map to decontaminate our results or modifying the dust law within the uncertainty range led to only small changes. In future work, we could try to test drastically different dust maps -- but so far, our footprint selection criteria appear robust.

For template deprojection, we tested the impact of using all vs. one contaminant template when deprojecting, and found it to not have significant impacts. As mentioned earlier, this indicates that we are working with data that is not heavily contaminated by observational systematics -- although this could be an artifact of using simulated skymaps rather than fully realistic object catalogs. Observational systematics are a modest limitation to the current state of the art, requiring e.g., discarding higher redshift bins (see e.g., \citealt{rodriguez+2022}). One simple(r) test for future work would be to try LSST survey simulations that have particularly significant observational impacts, e.g., bad seeing (potential issue), survey non-uniformity arising from fixed telescope pointing (for proof of principle since this is a resolved issue following \citealt{awan+2016}) or rolling cadences (an outstanding issue and a focus of research and development in DESC; \citealt{lochner+2022}). Another avenue would be to use realistic image simulations, e.g., like the DESC Data Challenge 2 \citealt{dc2}, even with the smaller footprint compared to the full LSST footprint, which we also leave for future work.

Finally, we tried using an incomplete covariance matrix instead of the full one, and we did not find it to have drastic impacts on our posteriors.

There are additional systematics that we have not tested, some of which are under study e.g., in DESC. These include imaging systematics mentioned earlier, which include impacts of colors and sample selection and deblending (which affects clustering measurements via e.g., star vs. galaxy misclassification), galaxy bias (see e.g., \citealt{nicola+2023}), and stellar contamination. We also note that we operated in the framework presented in \citetalias{desc-srd2018}, which may be idealistic, e.g., with 10 fixed-width redshift bins spanning the entire redshift range accessible with LSST. There are ways to optimize the sample selection, e.g., via redshift bins and $N(z)$s (see e.g., \citealt{tomochallenge,moskowitz+2023, euclid2021, rau+2024}) that we did not incorporate; we leave these investigations for future work.

Another aspect of the planned framework that could be interesting to explore is the number of $\ell$ samples; we have used 20 $\ell$-bins, which, although it speeds up the inference analysis via sparser sampling of the power spectra, can reduce the S/N.

Note also that we looked at each systematic independently -- in a real analysis, all the systematics will need to be confronted at the same time, not to mention in a joint-probes analysis framework such as with cosmic shear\footnote{Note that 1) a joint galaxy clustering and cosmic shear analysis would also enable measuring parameters such as $\sigma_8$ as a 3x2pt analysis breaks the degeneracy between $\sigma_8$ and galaxy bias; 2) impacts of \pz\ systematics would be different for cosmic shear given that the lensing kernel is very broad compared to the width of typical tomographic bins used for galaxy clustering.}. Understanding the impact of each systematic on its own is a good first step, as we push the forefronts of the analysis techniques to analyze what will be the most complex astronomical data. These studies are imperative as we finalize preparations for the analysis of LSST data.

\section{Acknowledgements}
\enum{
\bl \textbf{DESC acknowledgements}:
This paper has undergone internal review by the LSST Dark Energy Science Collaboration with Ignacio Sevilla-Noarbe, An\^ze Slosar, and Joe Zuntz as the internal reviewers, who provided extensive feedback leading to major transformations of the work. The authors also thank the DESC LSS Working Group, and in particular, David Alonso, Andrina Nicola, and Boris Leistedt, for their helpful feedback.
\bl \textbf{Author contributions}:
HA designed the pipeline, wrote the code, performed the analysis, and wrote the paper. EG helped with the project design, advised on the analysis, and provided feedback on the text. JS helped with some of the code validation and analysis design, and provided feedback on the text. ISN reviewed the text and content.
\bl \textbf{Grants}: 
HA acknowledges support from the Leinweber Postdoctoral Research Fellowship and DOE grant DE-SC009193.  HA also thanks the LSSTC Data Science Fellowship Program, which is funded by LSSTC, NSF Cybertraining Grant \#1829740, the Brinson Foundation, and the Moore Foundation, as participation in the program has benefited this work. E.G. acknowledges support from the U.S. Department of Energy, Office of Science, Office of High Energy Physics Cosmic Frontier Research program under award DE-SC0010008. ISN would like to acknowledge the support of grant PGC2018-094773-B-C33 of the Spanish Ministerio de Ciencia e Innovacion.
\bl \textbf{Compute resources}:
Initial developments of the analysis used the National Energy Research Scientific Computing Center (NERSC), a U.S. Department of Energy Office of Science User Facility located at Lawrence Berkeley National Laboratory, operated under Contract No. DE-AC02-05CH11231 using NERSC award HEP-ERCAP0022779. Shortly after the initial developments, we switched to the Caliburn supercomputer for the bulk of the analysis; Caliburn was supported by Rutgers and the State of New Jersey. Once Caliburn was decommissioned, we switched to the Amarel cluster, access to which is provided by the Office of Advanced Research Computing (OARC) at Rutgers, The State University of New Jersey. 

The DESC acknowledges ongoing support from the Institut National de Physique Nucléaire et de Physique des Particules in France; the Science \& Technology Facilities Council in the United Kingdom; and the Department of Energy, the National Science Foundation, and the LSST Corporation in the United States. DESC uses resources of the IN2P3 Computing Center (CC-IN2P3–Lyon/Villeurbanne—France) funded by the Centre National de la Recherche Scientifique; the National Energy Research Scientific Computing Center, a DOE Office of Science User Facility supported by the Office of Science of the U.S. Department of Energy under contract No. DE-AC02- 05CH11231; STFC DiRAC HPC Facilities, funded by UK BEIS National E-infrastructure capital grants; and the UK particle physics grid, supported by the GridPP Collaboration. This work was performed in part under DOE contract DE- AC02-76SF00515.
\bl \textbf{Codes}: We acknowledge using \healpy\ \citep{zonca+2019} which wraps \healpix\ \citep{gorski+2005}, and \ttt{chainconsumer} \citep{chainconsumer}. We also extensively used core \ttt{python} packages such as \ttt{numpy} \citep{numpy}, \ttt{matplotlib} \citep{matplotlib}, and \ttt{scipy} \citep{scipy}. Other codes used are referenced within the text.
}

\bibliographystyle{JHEP}
\bibliography{references}
\newpage
\appendix
\section{Best Fit Plots - Continued\label{sec: bestfits}}
\begin{figure}[!htb]
	\centering
    \includegraphics[width=\linewidth, trim={5 5 5 5}, clip=true]{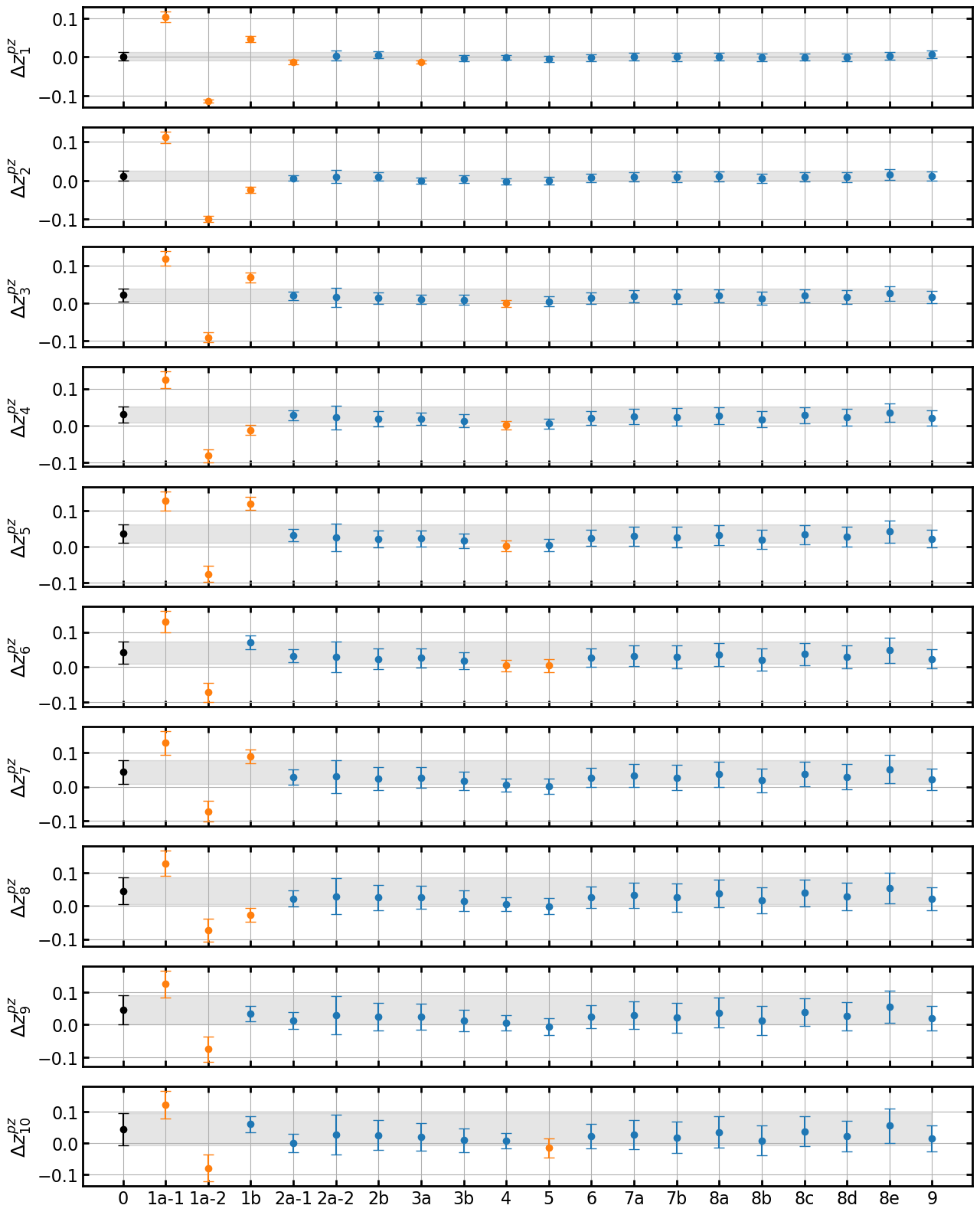}
	\caption{
    Best fit values for \pz\ shift parameters for all cases considered in this work; see \autoref{fig: bestfit cosmoparams} caption for more details.
    }
	\label{fig: bestfit delzs}
\end{figure}

\begin{figure}[!htb]
	\centering
	\includegraphics[width=\linewidth, trim={5 5 5 5}, clip=true]{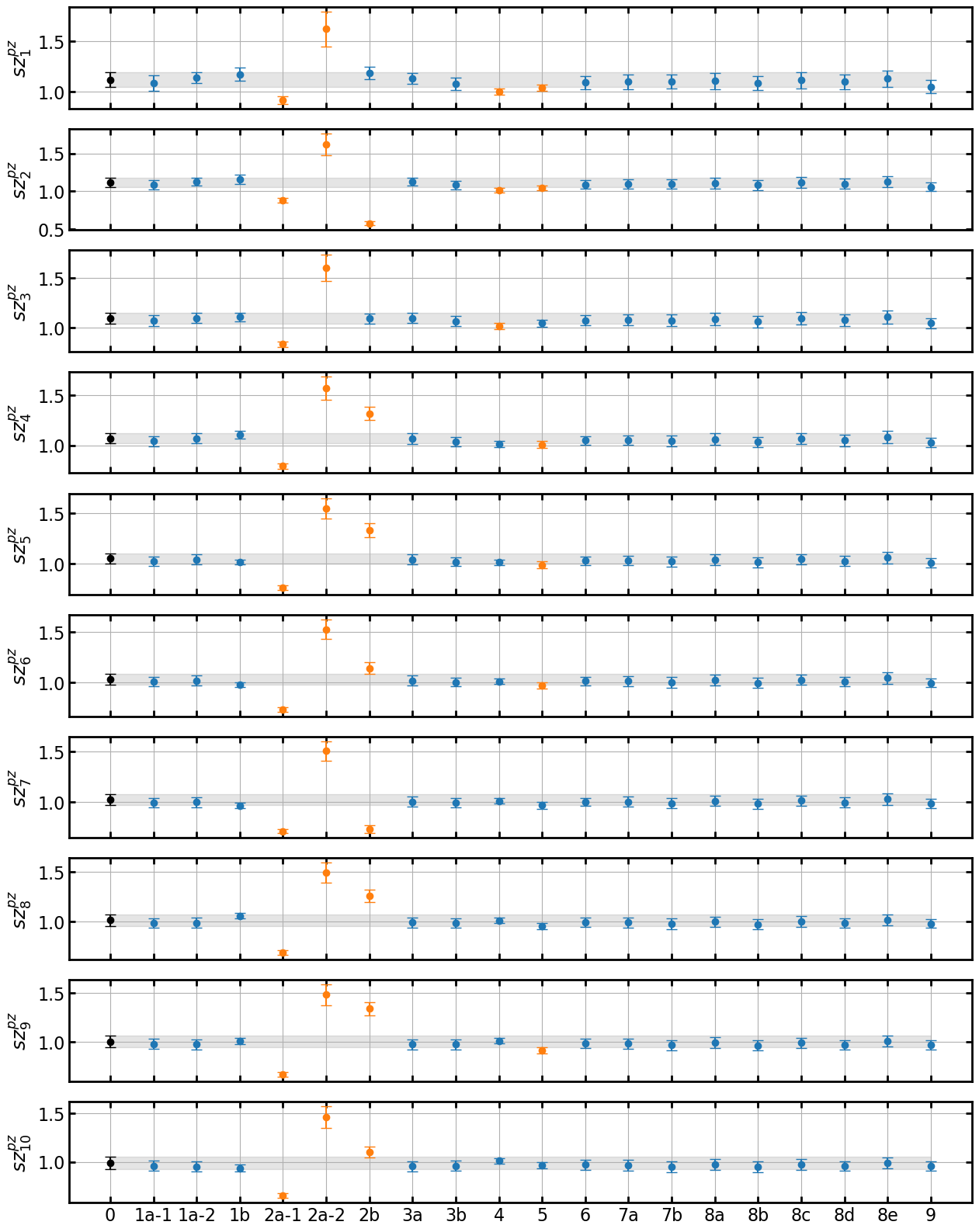}
	\caption{
    Best fit values for \pz\ stretch parameters for all cases considered in this work; see \autoref{fig: bestfit cosmoparams} caption for more details.
    }
	\label{fig: bestfit szs}
\end{figure}

\FloatBarrier
\section{Posterior Plots\label{sec: posteriors}}

\begin{landscape}
\begin{figure}[!htb]
    \vspace{-3em}
	\centering
	\includegraphics[width=0.7\paperwidth, trim={25 25 25 0}, clip=true]{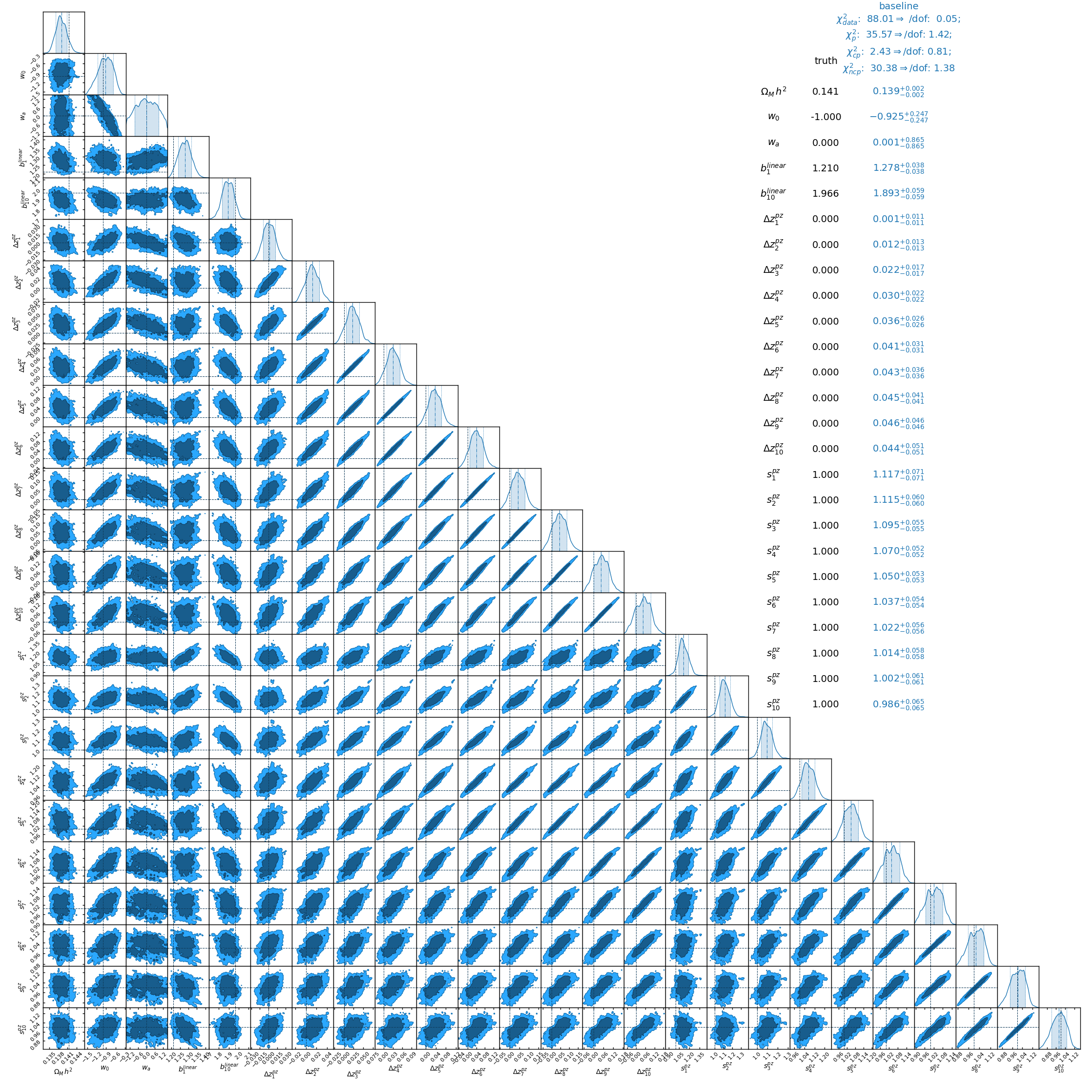}
	\caption{Baseline posteriors for all 25 parameters fit for in our inference analysis; subset shown in \autoref{fig: posteriors baseline}. As mentioned in the main text, while some of the best fit parameter values (in the legend table) are more than 1$\sigma$ away from the truth, the deviation is never more than 2$\sigma$; the degree of deviation is even less pronounced in the 2D plots. While the best fits are not perfect, they establish the baseline against which we compare the rest of the cases. See the caption of \autoref{tab: results1} for details regarding the various \chitwo\ statistics in the legend.
    }
	\label{fig: posteriors baseline full}
\end{figure}

\begin{figure}[!htb]
    \vspace{-3em}
	\centering
		\includegraphics[width=\paperwidth, trim={25 25 50 0}, clip=true]{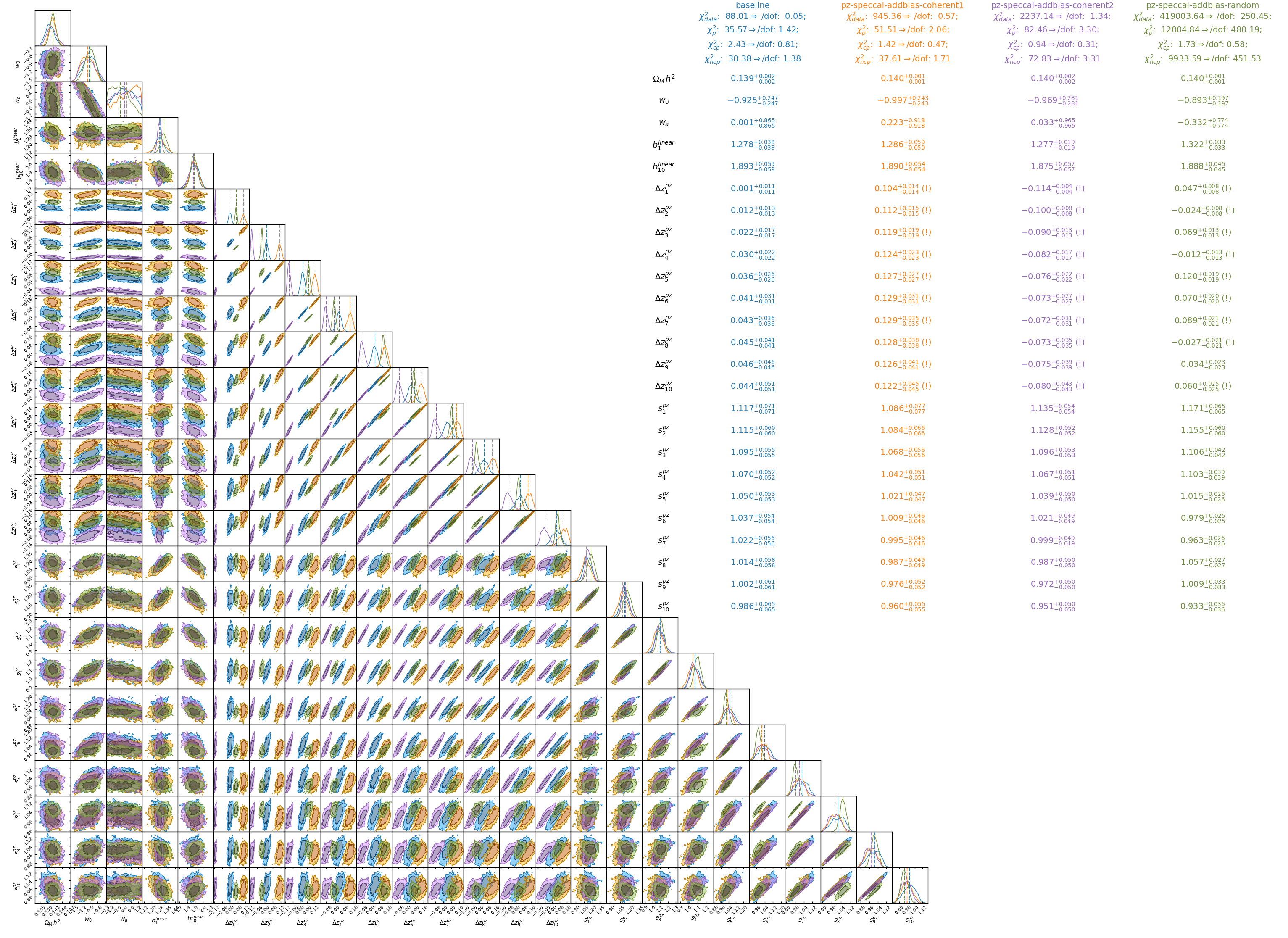}
	\caption{Posteriors for the case of coherent \pz\ spectroscopic calibration additive biases (with $\Delta_i \pm 0.1$) and random ones (chosen between $\pm$ 0.1), alongside those for \baseline\ best fit. In the legend, we show the best fit values as well as various statistics, as in \autoref{fig: posteriors baseline}; (!) indicates a $>1\sigma$ bias from \baseline. We see that this systematic has adverse, noticeable impacts, with the random biases causing the most damage.
	}
	\label{fig: posteriors pz shift}
\end{figure}

\begin{figure}[!htb]
    \vspace{-3em}
	\centering
	\includegraphics[width=\paperwidth, trim={25 25 50 0}, clip=true]{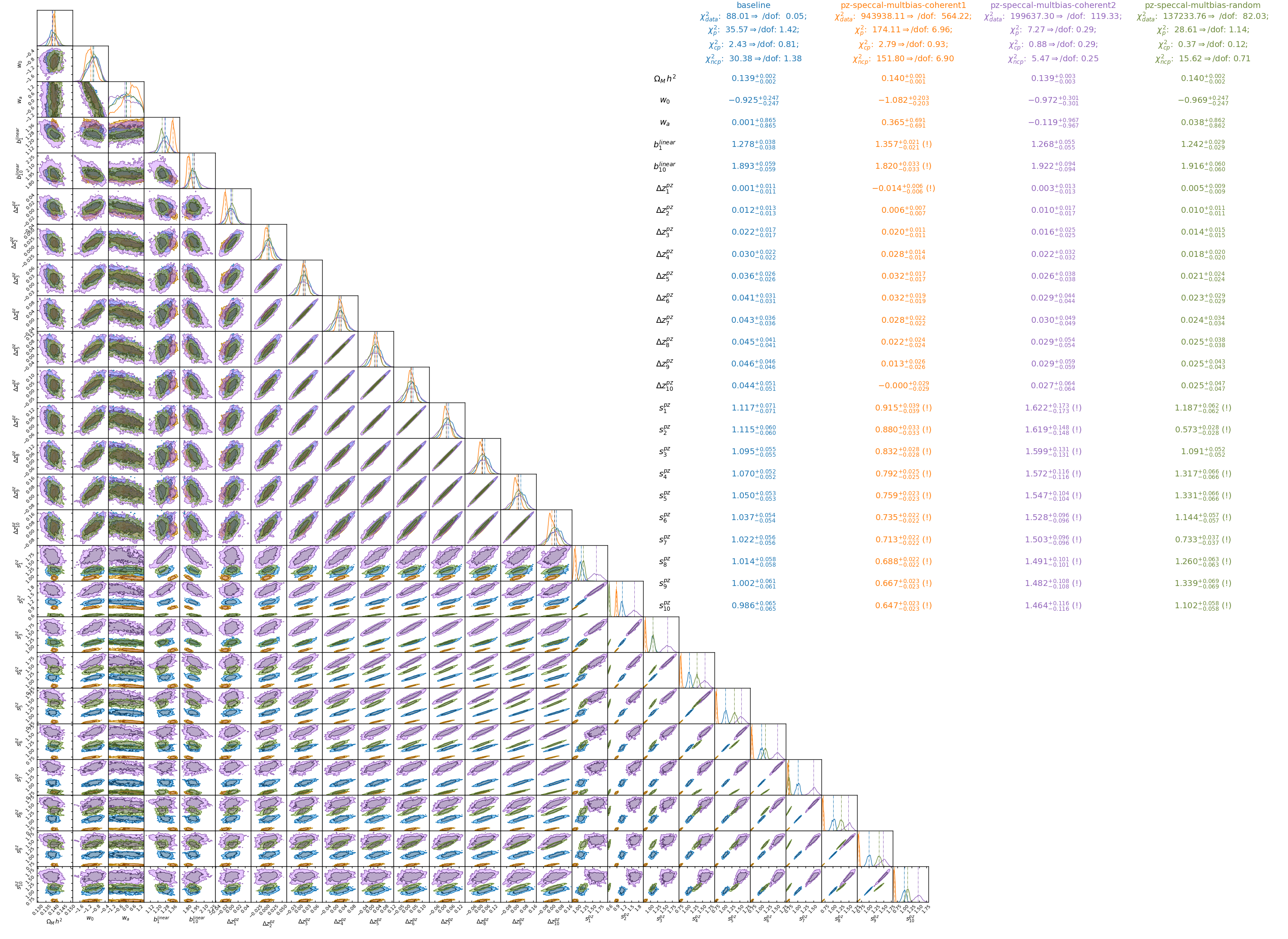}
	\caption{Posteriors for the case of coherent \pz\ spectroscopic calibration multiplicative biases (with $s_i = 0.75, 1.5$) and random ones (chosen between 0.5, 1.5), alongside those for \baseline. In the legend, we show the best fit values as well as various statistics, as in \autoref{fig: posteriors baseline}; (!) indicates a $>1\sigma$ bias from \baseline\ best fit. We see that all variations of this systematic lead to adverse, noticeable impacts.
 }
	\label{fig: posteriors pz stretch}
\end{figure}

\begin{figure}[!htb]
    \vspace{-3em}
	\centering
		\includegraphics[width=0.85\paperwidth, trim={25 25 25 0}, clip=true]{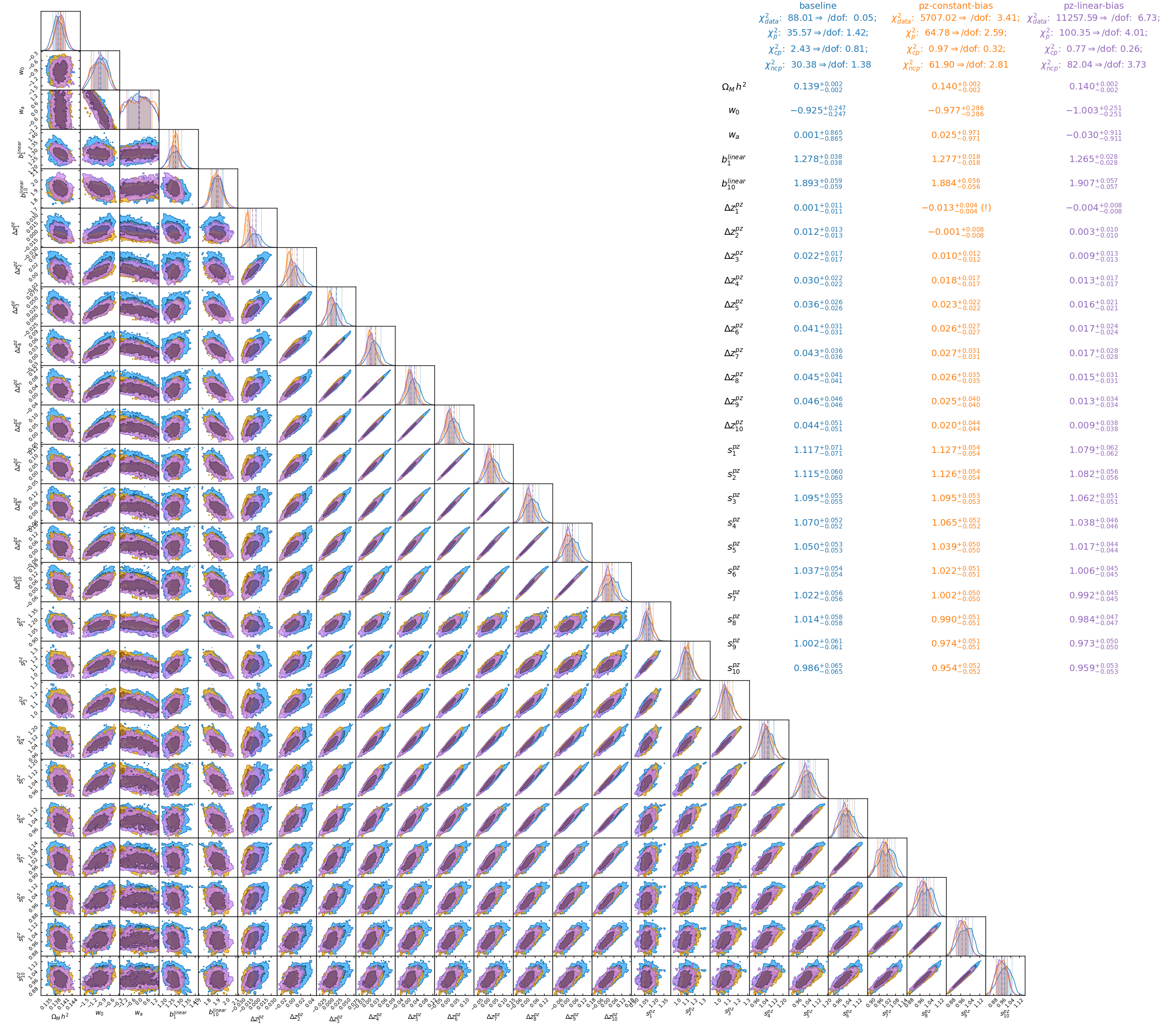}
	\caption{Posteriors for the case of constant \pz\ bias (maximum plausible, i.e., the width of the redshift bin, 0.1) and linear \pz\ bias ($0.1z$). In the legend, we show the best fit values as well as various statistics, as in \autoref{fig: posteriors baseline}; (!) indicates a $>1\sigma$ bias from \baseline\ best fit. We see that this systematic causes only some best fits to be biased, although the \chitwo\ statistics suffer a little. 
 }
	\label{fig: posteriors pz bias}
\end{figure}

\begin{figure}[!htb]
    \vspace{-3em}
	\centering
		\includegraphics[width=0.75\paperwidth, trim={25 25 25 0}, clip=true]{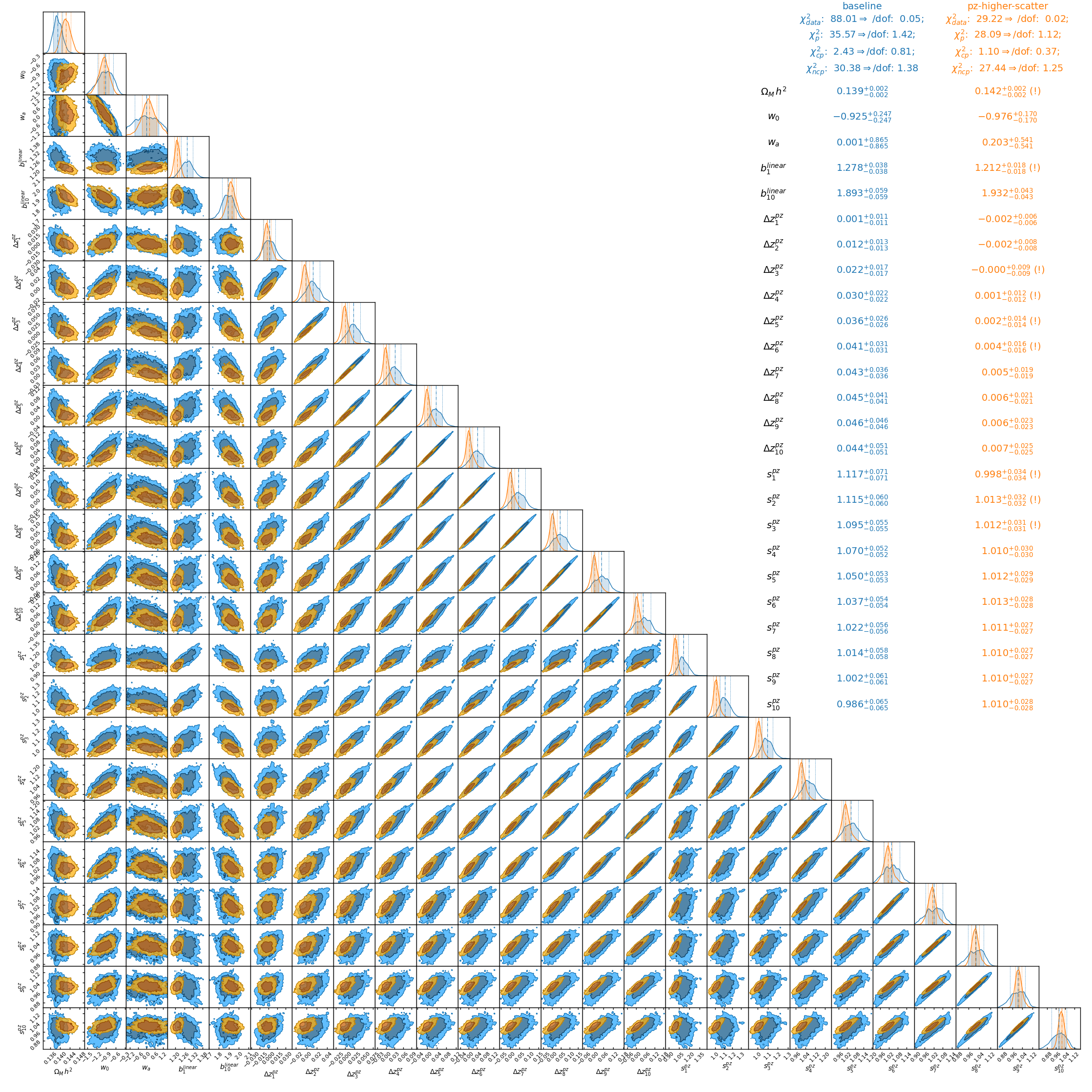}
	\caption{Posteriors for the case of larger \pz\ scatter amplitude (twice the baseline value; $\sigma_{z_0} = 0.06$). In the legend, we show the best fit values as well as various statistics, as in \autoref{fig: posteriors baseline}; (!) indicates a $>1\sigma$ bias from \baseline\ best fit. We see that while the best fits are biased, the \chitwo\ statistics do not worsen, indicating that this systematic can lead to trouble without giving an explicit indication (via \chitwodata).
    }
	\label{fig: posteriors pz large-scatter}
\end{figure}

\begin{figure}[!htb]
    \vspace{-3em}
		\centering
		\includegraphics[width=0.75\paperwidth, trim={25 25 25 0}, clip=true]{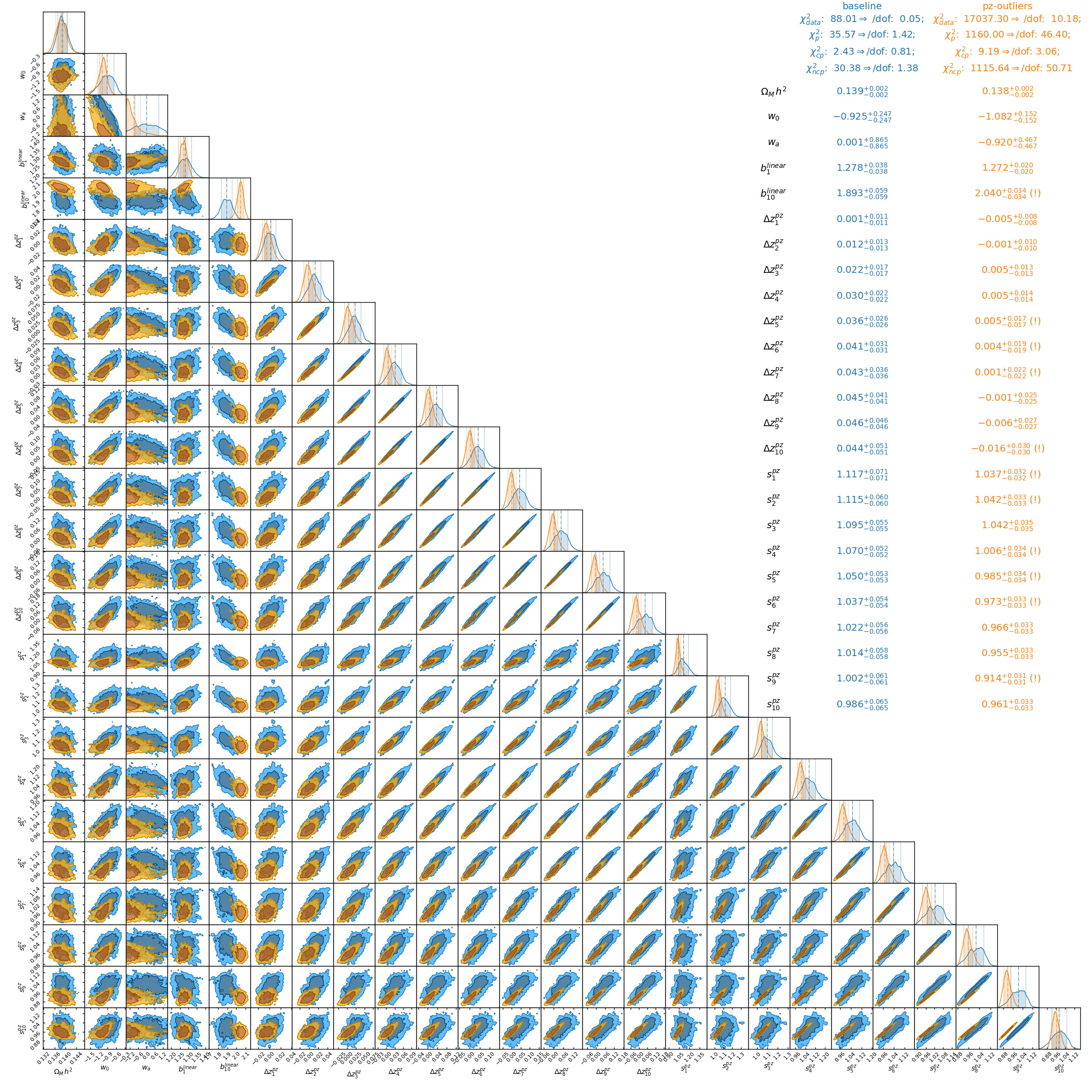}
	\caption{Posteriors for the case of 10\% \pz\ outliers. In the legend, we show the best fit values as well as various statistics, as in \autoref{fig: posteriors baseline}; (!) indicates a $>1\sigma$ bias from \baseline\ best fit. We see that this systematic has modest adverse impact, noticeable also via the \chitwo\ statistics.
    }
	\label{fig: posteriors pz outliers}
\end{figure}

\begin{figure}[!htb]
    \vspace{-3em}
	\centering
	\includegraphics[width=0.8\paperwidth, trim={25 25 25 0}, clip=true]{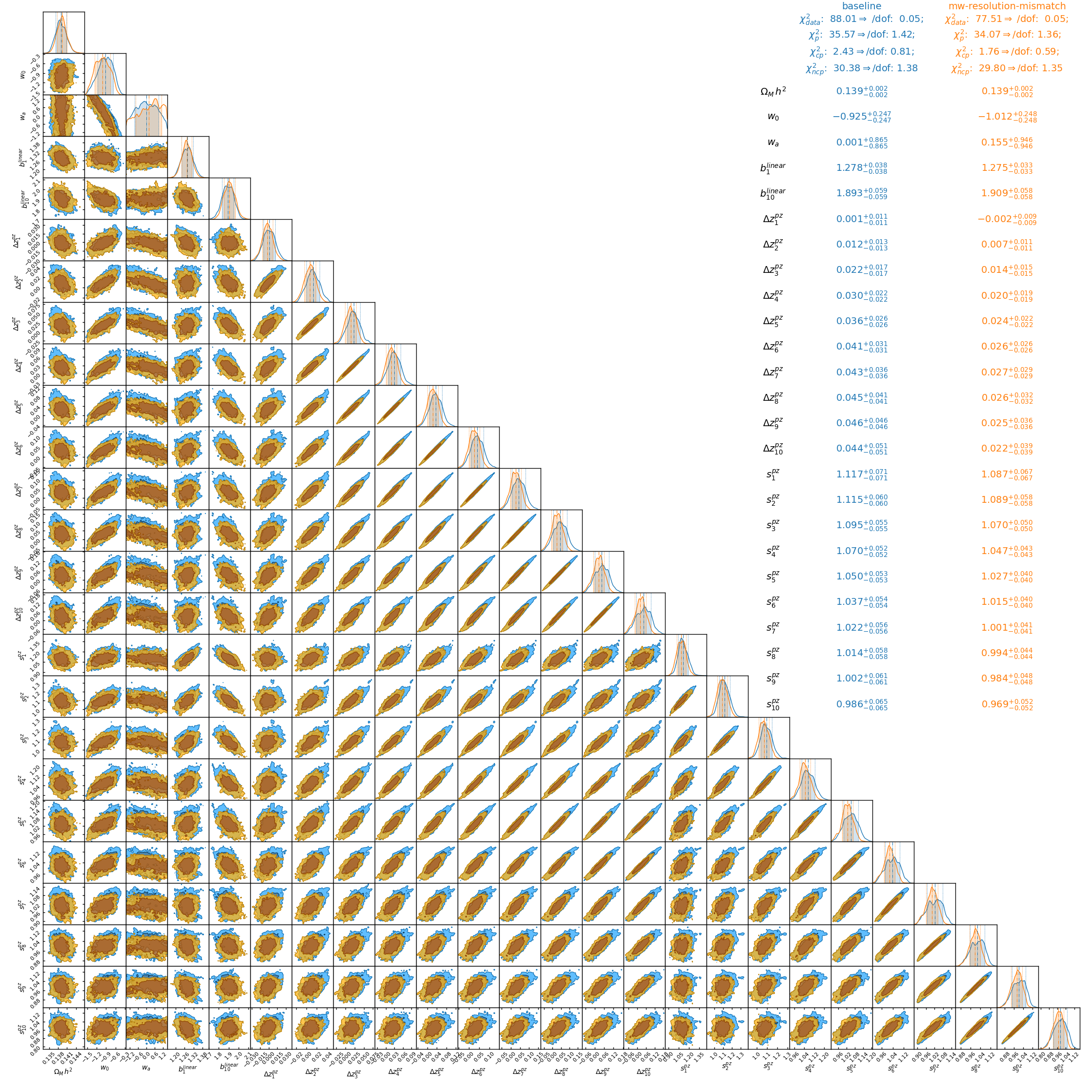}
	\caption{Posteriors for the case of resolution mismatch between the dust map used for generating mocks vs. that used in deprojection. In the legend, we show the best fit values as well as various statistics, as in \autoref{fig: posteriors baseline}; (!) indicates a $>1\sigma$ bias from \baseline\ best fit. We see that the mismatch does not lead to any significant differences compared to \baseline.}
	\label{fig: posteriors mw res-mismatch}
\end{figure}

\begin{figure}[!htb]
    \vspace{-3em}
	\centering
		\includegraphics[width=0.8\paperwidth, trim={25 25 25 0}, clip=true]{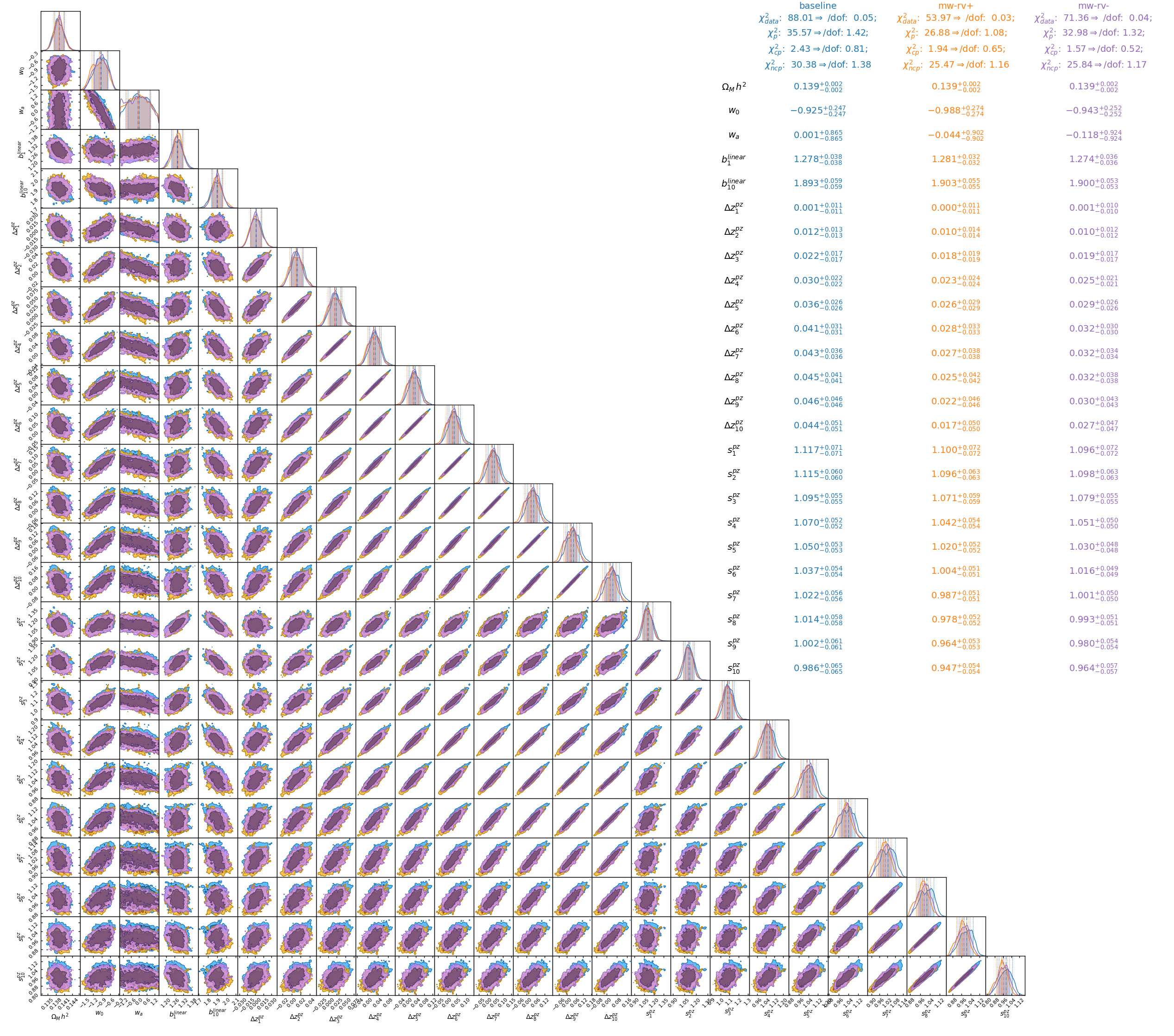}
	\caption{Posteriors for the case with a modified dust law (with $R_V = 3.1 \pm 1.8$ as opposed to 3.1 for \baseline). In the legend, we show the best fit values as well as various statistics, as in \autoref{fig: posteriors baseline}; (!) indicates a $>1\sigma$ bias from \baseline\ best fit. We see that none of the cases lead to significant differences compared to \baseline.}
	\label{fig: posteriors mw rv}
\end{figure}

\begin{figure}[!htb]
    \vspace{-3em}
	\centering
		\includegraphics[width=\paperwidth, trim={25 25 50 0}, clip=true]{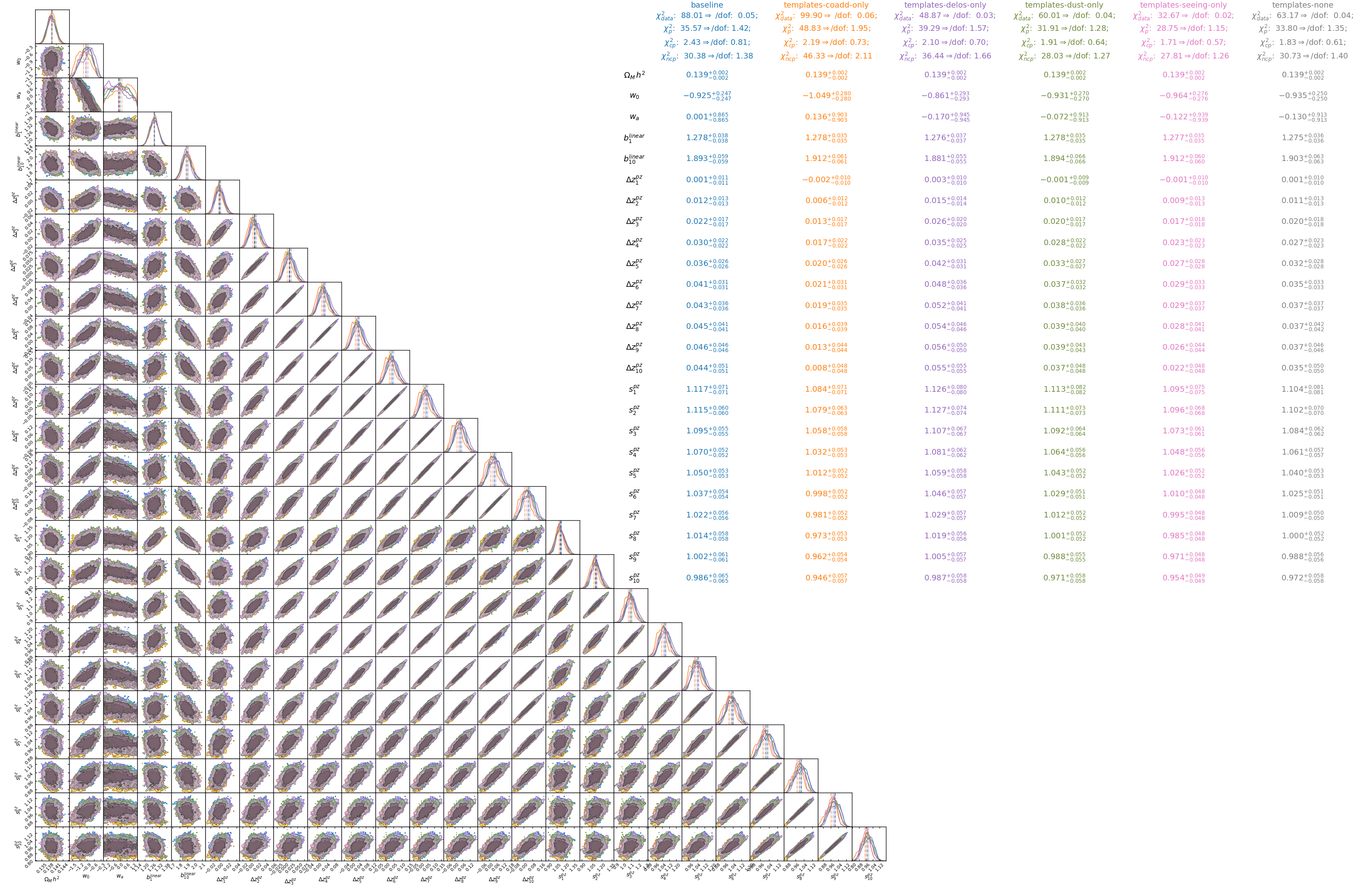}
    \caption{Posteriors for the case with no templates used for deprojection as well as the case that uses one of four templates; to clarify, 1+5 posteriors are shown, with legend entries for each. In the legend, we show the best fit values as well as various statistics, as in \autoref{fig: posteriors baseline}; (!) indicates a $>1\sigma$ bias from \baseline\ best fit. We see that none of the cases lead to significant differences compared to \baseline.}
	\label{fig: posteriors cts}
\end{figure}

\begin{figure}[!htb]
    \vspace{-3em}
	\centering
		\includegraphics[width=0.75\paperwidth, trim={25 25 25 0}, clip=true]{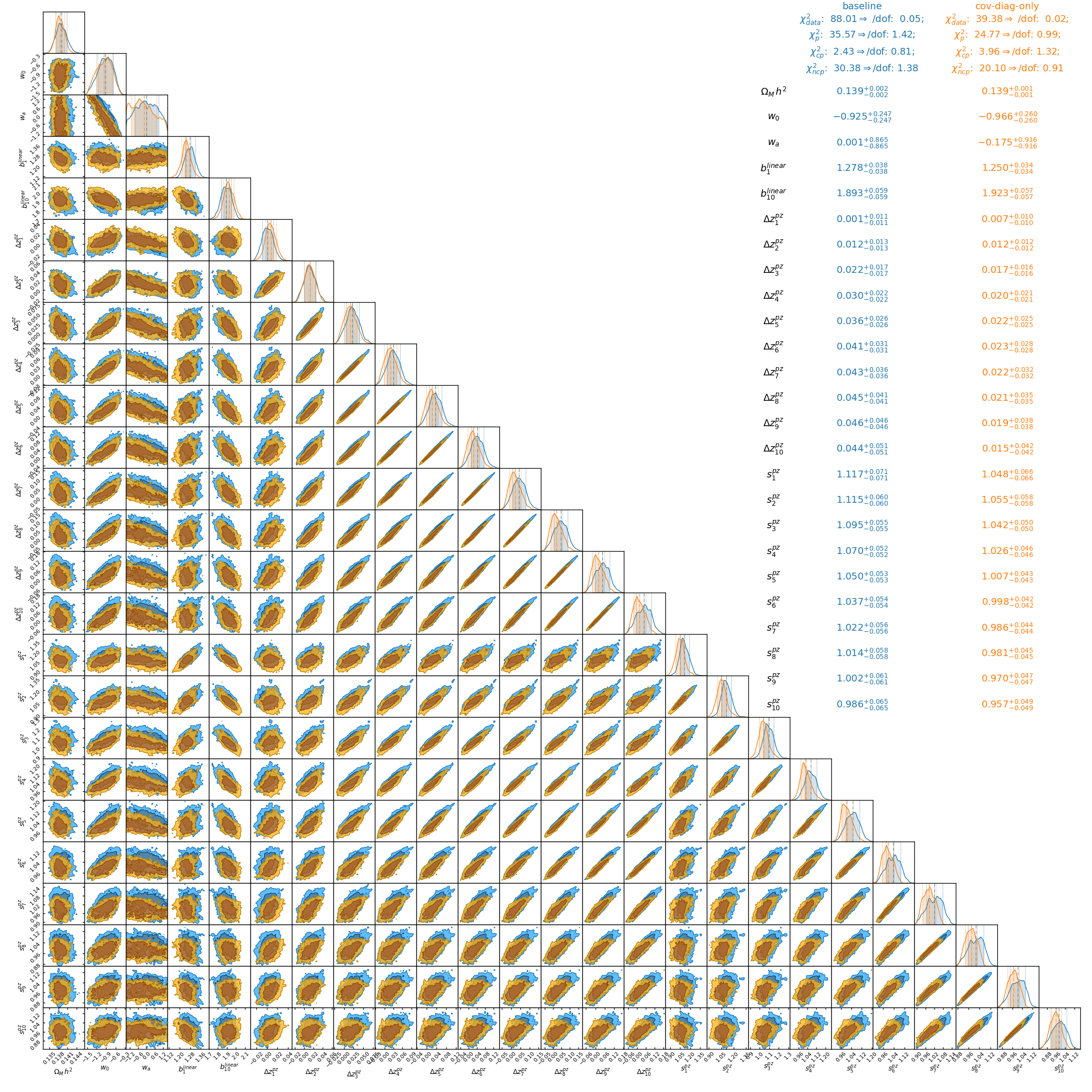}
	\caption{Posteriors for the case where we use only the diagonal of the covariance matrix in our likelihood analysis. In the legend, we show the best fit values as well as various statistics, as in \autoref{fig: posteriors baseline}; (!) indicates a $>1\sigma$ bias from \baseline\ best fit. We see that using only partial information does not lead to any significant differences compared to \baseline.}
	\label{fig: posteriors cov diag}
\end{figure}

\end{landscape}

\section{Impact of Poisson Noise\label{sec: theory-seeds}}
\begin{figure}[!htb]
	\centering
		\includegraphics[width=\linewidth, trim={5 5 5 5}, clip=false]{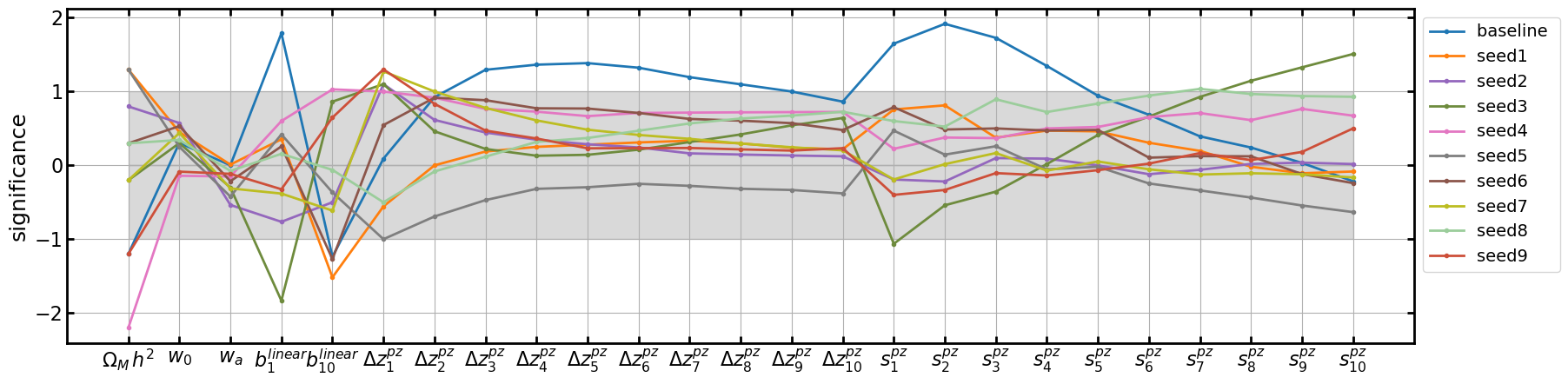}
	\caption{Best fits from 10 different variations of \baseline, each using a different seed. Significance here is defined as (best fit - truth) / $\sigma$ where $\sigma$ is the maximum between upper and lower best fit error. Among the 10 different seeds, we see up to 2$\sigma$ deviations from the truth - and that, by sheer coincidence, our \baseline\ seed is one of the worst.}
	\label{fig: theory-seeds bestfits}
\end{figure}

\begin{landscape}
\begin{figure}[!htb]
    \vspace{-3em}
	\centering
		\includegraphics[width=0.8\paperwidth, trim={25 25 25 0}, clip=true]{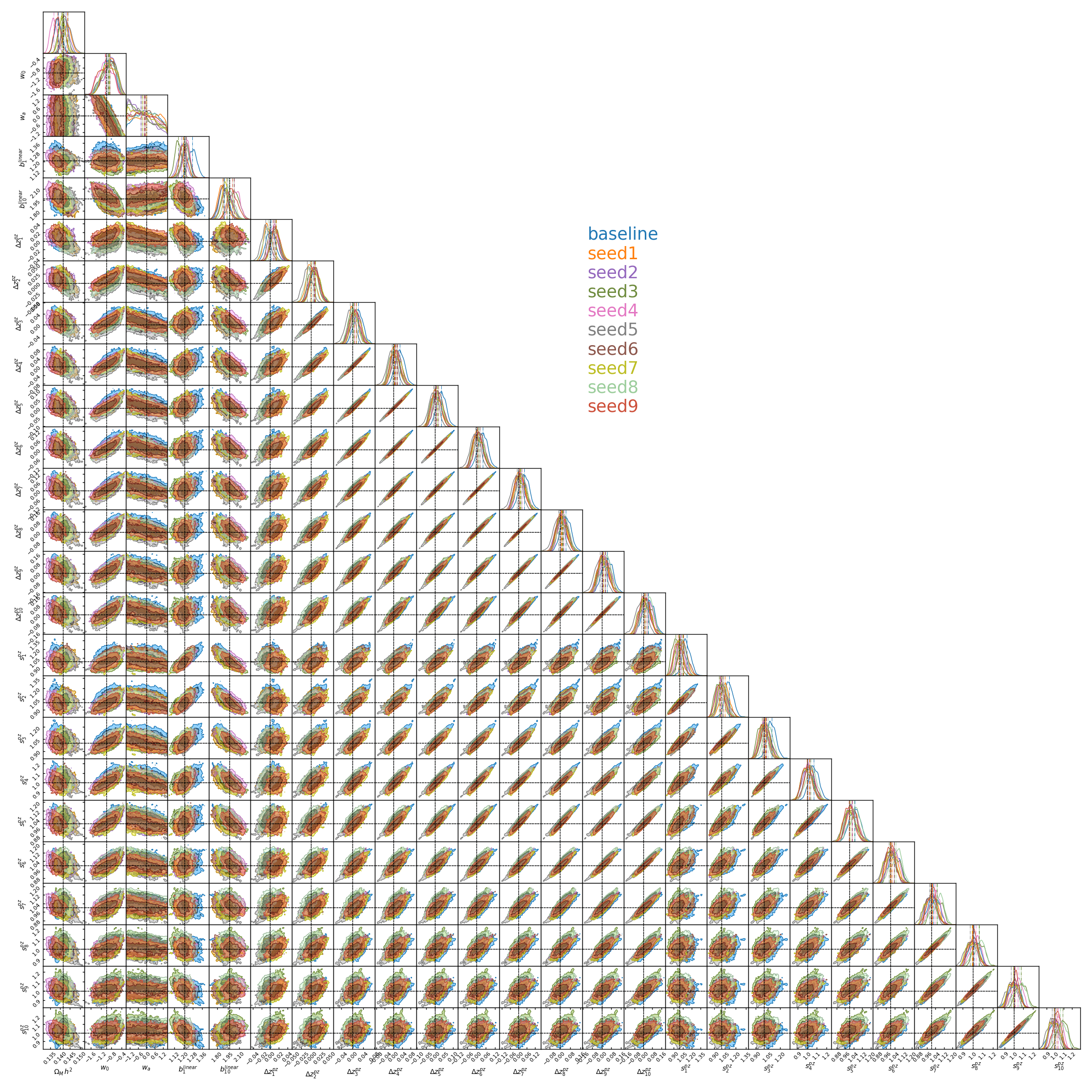}
	\caption{Posteriors for 10 different variations of \baseline, each using a different seed. We see that the posteriors are quite comparable to each other, although with 1-2$\sigma$ differences visible e.g., in the projected histograms. Black dotted lines show the truth, same in as \autoref{fig: posteriors baseline}.}
	\label{fig: posteriors theory-seeds}
\end{figure}
\end{landscape}

Since we realize our theory maps using a random seed, as explained in \autoref{sec: map generation}, we explore the impact of this randomness on our posteriors by creating 10 different variations of \baseline, each using a different seed. As mentioned in \autoref{sec: results}, we find that the specific seed can lead to 1-2$\sigma$ variation; this is visible in \autoref{fig: theory-seeds bestfits} where we plot the significance of each of the best fit parameters for all 10 cases. We also show the posteriors in \autoref{fig: posteriors theory-seeds}. Note that this should not impact our conclusions since we fix the seed for all analysis runs and only do a comparative analysis (against \baseline).

\end{document}

%% file: table2.tex
\newcolumntype{x}{>{\centering\arraybackslash}m{1cm}}
\newcolumntype{y}{>{\centering\arraybackslash}m{1.75cm}}
\newcolumntype{w}{>{\centering\arraybackslash}m{1.9cm}}
\newcolumntype{v}{>{\centering\arraybackslash}m{5cm}}
\newcolumntype{z}{>{\centering\arraybackslash}m{5cm}}
\newcolumntype{a}{>{\centering\arraybackslash}m{2.5cm}}

\let\oldsub=\subsectionautorefname
\let\oldsubsub=\subsubsectionautorefname

\makeatletter
\renewcommand*\subsubsectionautorefname{\@gobble}
\renewcommand*\subsectionautorefname{\@gobble}
\makeatother

\begin{table*}[!htb]
\resizebox{\linewidth}{!}{
\begin{tabular}{
|x|y!{\vrule width 1.2pt}
z|w!{\vrule width 1.2pt}
v!{\vrule width 1.2pt}
a|y|
}
\Xhline{3\arrayrulewidth}
\multicolumn{2}{|c!{\vrule width 1.2pt}}{} &
\multicolumn{2}{c!{\vrule width 1.2pt}}{Construction of Density Maps} &
Inputs to \nmt &
\multicolumn{2}{c|}{Likelihood Analysis}  \\
\hline\hline
\# (section) & Systematic in Test &
\pz\ model via $n_i(z)$ & MW impact &
template maps &
initial \pz\ model via $n_i(z)$ & covariance \\
\Xhline{3\arrayrulewidth}
0 (\autoref{sec: pz baseline})
&
NONE
&
\baseline\ ($\delta_z$=0, $\sigma_{z_0}$=0.03, $\Delta_i$=0, $s_i$=1, $f_\mathrm{outliers}$=0)  & baseline ($R_V$=3.1; $N_\mathrm{side}$ 1024)
&
(map 1) 5$\sigma$ coadded depth with baseline MW dust extinction, \newline
(map 2) seeing map, \newline
(map 3) $\delta_\mathrm{OS}^\mathrm{no\ dust}$ map
&
baseline
&
full \\
\hline
1a (\autoref{sec: pz shift})
& \pz &
\textbf{spec-calibration additive bias} (coherent: same, non-zero $\Delta_i$ for all redshift bins) & baseline &
maps 1-3
&
\textbf{baseline} & full \\
\hline
1b (\autoref{sec: pz shift})
& \pz &
\textbf{spec-calibration additive bias} (random: different, random $\Delta_i$ for each redshift bin) & baseline &
maps 1-3
&
\textbf{baseline} & full \\
\hline
2a (\autoref{sec: pz stretch})
& \pz &
\textbf{spec-calibration multiplicative bias} (coherent: same, $\neq$ 1 $s_i$ for all redshift bins) & baseline & 
maps 1-3
&
\textbf{baseline} & full \\
\hline
2b (\autoref{sec: pz stretch})
& \pz &
\textbf{spec-calibration multiplicative bias} (random: different, random $s_i$ for each redshift bin) & baseline & 
maps 1-3
&
\textbf{baseline} & full \\
\hline
3a (\autoref{sec: pz bias})
& \pz &
\textbf{biased} (constant: $\delta_z$ nonzero but constant) & baseline &
maps 1-3
&
 \textbf{biased} & full \\
\hline
3b (\autoref{sec: pz bias})
& \pz &
\textbf{biased} (linear: $\delta_z \propto z$) & baseline & 
maps 1-3
&
 \textbf{biased} & full \\
\hline
4 (\autoref{sec: pz scatter})
& \pz &
\textbf{(more) uncertain \pz} ($\sigma_{z_0}$ higher) & baseline & 
maps 1-3
&
 \textbf{(more) uncertain \pz} & full \\
\hline
5 (\autoref{sec: pz outliers})
& \pz &
\textbf{\pz\ outliers} ($f_\mathrm{outliers}$ nonzero) & baseline & 
maps 1-3
&
\textbf{\pz\ outliers} & full \\
\hline
6 (\autoref{sec: mw res-mismatch})
& dust &
baseline & \textbf{baseline} & 
5$\sigma$ coadded depth \textbf{with a lower resolution ($N_\mathrm{side} \ 64$) dust extinction map}, \newline
maps 2-3
&
baseline & full \\
\hline
7a (\autoref{sec: mw rv change})
& dust &
baseline & \textbf{different dust law} ($R_V^\mathrm{base} +$ 0.18) & 
5$\sigma$ coadded depth with \textbf{baseline dust extinction}, \newline
maps 2-3
&
baseline & full  \\
\hline
7b (\autoref{sec: mw rv change})
& dust &
baseline & \textbf{different dust law} ($R_V^\mathrm{base} -$ 0.18) & 
5$\sigma$ coadded depth with \textbf{baseline dust extinction}, \newline
maps 2-3
&
baseline & full  \\
\hline
8a (\autoref{sec: ct})
& templates &
baseline & baseline & 
None
&
baseline & full  \\
\hline
8b (\autoref{sec: ct})
& templates &
baseline & baseline & 
map 1
&
baseline & full  \\
\hline
8c (\autoref{sec: ct})
& templates &
baseline & baseline & 
map 2
&
baseline & full  \\
\hline
8d (\autoref{sec: ct})
& templates &
baseline & baseline & 
map 3
&
baseline & full  \\
\hline
8e (\autoref{sec: ct})
& templates &
baseline & baseline & 
baseline dust extinction map
&
baseline & full  \\
\hline
9  (\autoref{sec: cov}) & covariance &
baseline & baseline & 
maps 1-3
&
baseline & \textbf{diagonal only} \\
\hline
\end{tabular}
}
\let\subsectionautorefname=\oldsub
\let\subsubsectionautorefname=\oldsubsub
\caption{List of the various cases studied. Cases 1-9 differ from \baseline\ by only one variable (bold entries in each row highlight the mismatch, where applicable, between the underlying truth and what's assumed in the likelihood analysis) -- to ensure that we test the impact of each change independently. In all cases, the parameters fitted via inference include 3 cosmology parameters, 2 galaxy bias parameters, and one shift and one stretch parameter for each redshift bin, as explained in \autoref{sec: estimation of params}. \label{tab: cases}}
\end{table*}
\let\subsectionautorefname=\oldsub
\let\subsubsectionautorefname=\oldsubsub

%% file: table3.tex
\begin{table}[!htb]
\centering
\renewcommand{\arraystretch}{1.5}
\hspace*{-4em}
\resizebox{1.2\paperwidth}{!}{
\begin{tabular}{c | c |  c | c | c | c | c | c | c | c | c | c | c | c | c | c | c | c | c | c | c | c | c | c | c | c | c | c | c | c | c  }
\# & case &  $\Omega_M\,h^2$ &  $w_0$ &  $w_a$ &  $b_{1}^{linear}$ &  $b_{10}^{linear}$ & $\chi^2_{data}$ / dof & $\chi^2_{params}$ /dof & $\chi^2_{cp}$ / dof & $\chi^2_{ncp}$ / dof \\ 
 \hline \hline
0 & baseline & $0.139_{-0.002}^{+0.002}$ & $-0.925_{-0.247}^{+0.247}$ & $0.001_{-0.865}^{+0.865}$ & $1.278_{-0.038}^{+0.038}$ & $1.893_{-0.059}^{+0.059}$  & 0.05 & 1.42 & 0.81 & 1.38  \\
\hline    
1a & coherent spec-calibration additive bias 1 (\autoref{sec: pz shift}) & $0.140_{-0.001}^{+0.001}$ & $-0.997_{-0.243}^{+0.243}$ & $0.223_{-0.918}^{+0.918}$ & $1.286_{-0.050}^{+0.050}$ & $1.890_{-0.054}^{+0.054}$  & 0.57 & 2.06 & 0.47 & 1.71  \\
\hline    
& coherent spec-calibration additive bias 2 (\autoref{sec: pz shift}) & $0.140_{-0.002}^{+0.002}$ & $-0.969_{-0.281}^{+0.281}$ & $0.033_{-0.965}^{+0.965}$ & $1.277_{-0.019}^{+0.019}$ & $1.875_{-0.057}^{+0.057}$ & 1.34 & 3.30 & 0.31 & 3.31  \\
\hline    
1b & random spec-calibration additive bias (\autoref{sec: pz shift}) & $0.140_{-0.001}^{+0.001}$ & $-0.893_{-0.197}^{+0.197}$ & $-0.332_{-0.774}^{+0.774}$ & $1.322_{-0.033}^{+0.033}$ & $1.888_{-0.045}^{+0.045}$  & 250.45 & 480.19 & 0.58 & 451.53  \\
\hline    
2a & coherent spec-calibration multiplicative bias 1 (\autoref{sec: pz stretch}) & $0.140_{-0.001}^{+0.001}$ & $-1.082_{-0.203}^{+0.203}$ & $0.365_{-0.691}^{+0.691}$ & $1.357_{-0.021}^{+0.021}$ (!) & $1.820_{-0.033}^{+0.033}$ (!)  & 564.22 & 6.96 & 0.93 & 6.90  \\
\hline    
& coherent spec-calibration multiplicative bias 2 (\autoref{sec: pz stretch}) & $0.139_{-0.003}^{+0.003}$ & $-0.972_{-0.301}^{+0.301}$ & $-0.119_{-0.967}^{+0.967}$ & $1.268_{-0.055}^{+0.055}$ & $1.922_{-0.094}^{+0.094}$ & 119.33 & 0.29 & 0.29 & 0.25  \\
\hline    
2b & random spec-calibration multiplicative bias (\autoref{sec: pz stretch}) & $0.140_{-0.002}^{+0.002}$ & $-0.969_{-0.247}^{+0.247}$ & $0.038_{-0.862}^{+0.862}$ & $1.242_{-0.029}^{+0.029}$ & $1.916_{-0.060}^{+0.060}$  & 82.03 & 1.14 & 0.12 & 0.71  \\
\hline    
3a & constant \pz\ bias (\autoref{sec: pz bias}) & $0.140_{-0.002}^{+0.002}$ & $-0.977_{-0.286}^{+0.286}$ & $0.025_{-0.971}^{+0.971}$ & $1.277_{-0.018}^{+0.018}$ & $1.884_{-0.056}^{+0.056}$  & 3.41 & 2.59 & 0.32 & 2.81  \\ \hline
3b & linear \pz\ bias (\autoref{sec: pz bias}) & $0.140_{-0.002}^{+0.002}$ & $-1.003_{-0.251}^{+0.251}$ & $-0.030_{-0.911}^{+0.911}$ & $1.265_{-0.028}^{+0.028}$ & $1.907_{-0.057}^{+0.057}$  & 6.73 & 4.01 & 0.26 & 3.73  \\
\hline
4 & higher \pz\ uncertainty (\autoref{sec: pz scatter}) & $0.142_{-0.002}^{+0.002}$ (!) & $-0.976_{-0.170}^{+0.170}$ & $0.203_{-0.541}^{+0.541}$ & $1.212_{-0.018}^{+0.018}$ (!) & $1.932_{-0.043}^{+0.043}$  & 0.02 & 1.12 & 0.37 & 1.25  \\
\hline
5 & \pz\ outliers (\autoref{sec: pz outliers}) & $0.138_{-0.002}^{+0.002}$ & $-1.082_{-0.152}^{+0.152}$ & $-0.920_{-0.467}^{+0.467}$ & $1.272_{-0.020}^{+0.020}$ & $2.040_{-0.034}^{+0.034}$ (!)   & 10.18 & 46.40 & 3.06 & 50.71  \\
\hline
6 & dust map resolution mismatch (\autoref{sec: mw res-mismatch}) & $0.139_{-0.002}^{+0.002}$ & $-1.012_{-0.248}^{+0.248}$ & $0.155_{-0.946}^{+0.946}$ & $1.275_{-0.033}^{+0.033}$ & $1.909_{-0.058}^{+0.058}$  & 0.05 & 1.36 & 0.59 & 1.35  \\
\hline    
7a & different dust law ($R_V$=2.92) (\autoref{sec: mw rv change}) & $0.139_{-0.002}^{+0.002}$ & $-0.943_{-0.252}^{+0.252}$ & $-0.118_{-0.924}^{+0.924}$ & $1.274_{-0.036}^{+0.036}$ & $1.900_{-0.053}^{+0.053}$  & 0.04 & 1.32 & 0.52 & 1.17  \\
\hline
7b & different dust law ($R_V$=3.28) (\autoref{sec: mw rv change}) & $0.139_{-0.002}^{+0.002}$ & $-0.988_{-0.274}^{+0.274}$ & $-0.044_{-0.902}^{+0.902}$ & $1.281_{-0.032}^{+0.032}$ & $1.903_{-0.055}^{+0.055}$  & 0.03 & 1.08 & 0.65 & 1.16  \\
\hline    
8a & no template (\autoref{sec: ct}) & $0.139_{-0.002}^{+0.002}$ & $-0.935_{-0.250}^{+0.250}$ & $-0.130_{-0.913}^{+0.913}$ & $1.275_{-0.036}^{+0.036}$ & $1.903_{-0.063}^{+0.063}$  & 0.04 & 1.35 & 0.61 & 1.40  \\
\hline    
8b & depth template only (\autoref{sec: ct}) & $0.139_{-0.002}^{+0.002}$ & $-1.049_{-0.280}^{+0.280}$ & $0.136_{-0.903}^{+0.903}$ & $1.278_{-0.035}^{+0.035}$ & $1.912_{-0.061}^{+0.061}$  & 0.06 & 1.95 & 0.73 & 2.11  \\ \hline    
8c & dust extinction template only (\autoref{sec: ct}) & $0.139_{-0.002}^{+0.002}$ & $-0.931_{-0.270}^{+0.270}$ & $-0.072_{-0.913}^{+0.913}$ & $1.278_{-0.035}^{+0.035}$ & $1.894_{-0.066}^{+0.066}$  & 0.04 & 1.28 & 0.64 & 1.27  \\
\hline    
8d & seeing template only (\autoref{sec: ct}) & $0.139_{-0.002}^{+0.002}$ & $-0.964_{-0.276}^{+0.276}$ & $-0.122_{-0.939}^{+0.939}$ & $1.277_{-0.035}^{+0.035}$ & $1.912_{-0.060}^{+0.060}$  & 0.02 & 1.15 & 0.57 & 1.26  \\
\hline    
8e & $\delta_\mathrm{OS}$ template only (\autoref{sec: ct}) & $0.139_{-0.002}^{+0.002}$ & $-0.861_{-0.293}^{+0.293}$ & $-0.170_{-0.945}^{+0.945}$ & $1.276_{-0.037}^{+0.037}$ & $1.881_{-0.055}^{+0.055}$  & 0.03 & 1.57 & 0.70 & 1.66  \\
\hline    
9 & incomplete covariance (\autoref{sec: cov}) & $0.139_{-0.001}^{+0.001}$ & 
$-0.966_{-0.260}^{+0.260}$ & $-0.175_{-0.916}^{+0.916}$ & $1.250_{-0.034}^{+0.034}$ & $1.923_{-0.057}^{+0.057}$  & 0.02 & 0.99 & 1.32 & 0.91  \\
\hline
\hline
\end{tabular}
}
\caption{Best fit data for some of the parameters, tabulated for each case considered; all the non-zero cases should be compared against the zeroth one, \baseline. The statistics included are:
1) \chitwodata\ which is what is effectively used in the inference (=$-$2log$\mathcal{L})$; 2) \chitwop\ which tests the best fit parameters against the truth while accounting for parameter covariance (from \chainconsumer); 3) \chitwocp\ which is the same as \chitwop\ but just for the cosmology parameters (first three); and finally, 4) \chitwoncp\ which is the same as \chitwop\ but just for the non-cosmology parameters. In no case except the last do we find \chitwocp\ to be large when \chitwodata\ and \chitwoncp\ are not, a reassuring result. Also, the best fit cosmology parameters seem robust in the face of most systematics. Note that the entries with (!) are the ones with $>1\sigma$ bias against \baseline, with bias defined as in \autoref{eq: bias def}.  \label{tab: results1}}
\end{table}